\documentstyle[prb,eqsecnum,aps,epsf]{revtex}

\font\bba=msbm10 scaled 1200
\font\bbb=msbm8 
\font\bbc=msbm6 
\newfam\bbfam
\textfont\bbfam=\bba
\scriptfont\bbfam=\bbb
\scriptscriptfont\bbfam=\bbc
\def\bb{\fam\bbfam\bba}
\def\R{{\bb R}}
\def\Z{{\bb Z}}

\begin{document}
\draft
\title
{ Statistical Field Theory for Simple fluids : Mean Field and Gaussian
Approximations}
\author{J. - M. Caillol \thanks{e-mail: Jean-Michel.Caillol@th.u-psud.fr}}
\address{LPT - CNRS (UMR 8627) \\
	 Bat. 210,
	 Universit\'e de Paris Sud \\
	 F-91405 Orsay Cedex, France}
         
\date{\today}
\maketitle
\begin{abstract}
We present an exact field theoretical representation of the 
statistical mechanics of simple classical liquids with short-ranged pairwise 
additive interactions. The action of the field theory is obtained 
by performing a Hubbard-Stratonovich transformation of the 
configurational Boltzmann factor.
The  mean field and Gaussian approximations of the theory are derived and 
applications to the liquid-vapour transition  considered. 
\end{abstract}

\section{Introduction}
This article on a somehow new presentation of the theory of liquids
is dedicated to \mbox{D. Levesque} at the occasion of his $65^{th}$ 
birthday.  Since Dominique and I share the same office, it was nearly 
impossible to keep secret the writing of the paper and I could not help 
myself discussing with him some parts of the manuscript.
His deep knowledge of the 
theory of liquids  made his remarks 
always pertinent and the final status of this 
article owes very much to our discussions. 

The purpose of this work is to try to apply some of the techniques of 
statistical field theory to the physics of simple  liquids. 
By "simple" we mean a liquid made of a single, chemically inert component 
which can be satisfactorily modelled by 
pairwise additive and spherically symmetric interactions\cite{Hansen}.
More 
precisely we shall consider the case of a pair potential $v(r)$ which 
can be written as the sum of a hard core interaction plus a short-ranged
tail $-w(r)$ which is supposed to decay faster than 
$1/r^{3+\epsilon}$ at large $r$ so that the thermodynamic limit 
exists \cite{Hansen}.
Long-ranged electrostatic interactions are thus 
excluded of the present study as well as molecular liquids or mixtures.
With the help of a 
Kac-Siegert-Stratonovich-Hubbard-Edwards (KSSHE) 
\cite{Kac,Siegert,Strato,Hubbard1,Edwards} transformation it 
is formally possible to reexpress the grand partition function of this 
system of decorated hard spheres as  the grand partition function of 
bare hard spheres in an external 
stochastic potential or "field" $\phi$ with a Gaussian measure
\cite{Siegert,Brilliantov,Brilliantov2,Orland,Cai-Raim}. We 
can thus describe  the fluid with the help of a field theory characterised by an
action ${\cal H}(\phi)$ which however is quite complicated since, 
besides a simple quadratic term, it also 
includes the grand potential of the hard spheres which is a non local 
functional of  the field $\phi$. Quite immodestly we however
assume that this functional is perfectly known.
Of course, it would be  more satisfactory to treat 
the hard sphere interactions 
at the same level as the tail $ w(r)$ but the KSSHE transform 
requires a
well behaved interaction (namely its Fourier transform must exist and 
be well behaved, see e.g. the appendix) and, unfortunately, hard core or
more realistic less singular
repulsive short-ranged potentials do 
not fulfill these mathematical requirements.

All the sophisticated techniques developped in statistical field theory 
\cite{Ma,Parisi,Zinn,Dowrick}
can thus be applied {\em a priori} to the case of liquids. Feynman 
graphs are no more difficult to compute than Mayer graphs which 
makes possible to perform effectively perturbative expansions of various 
thermodynamic quantities attached to the liquid in the KSSHE 
representation.  
Recently this type of approach was used 
to rediscover the low fugacity \cite{Orland,Cai-Raim} and high 
temperature \cite{Cai-Raim} expansions of the free energy and the
equation of state of the restricted 
primitive model of electrolytes (RPM), results which had been
obtained years ago before by awkward
graph  resummation techniques \cite{Mayer,Lebo}.

Another powerfull technique of statistical field theory 
is the so-called loop expansion which amounts to expand functionally 
the action ${\cal H}(\phi)$ around a saddle-point 
\cite{Parisi,Zinn,Dowrick}. 
The lowest order (zero loop) of the expansion,
defines the mean field  (MF) level of the theory. The MF theory of 
liquids in the KSSHE representation has been studied recently by 
Brilliantov et al.  both  for homogeneous liquids 
\cite{Brilliantov} 
and for liquids inhomogeneous in one direction of space
\cite{Brilliantov2}. In the former case, it yields 
to a van de Waals like theory of the liquid-vapour transition and, in 
the latter one, it allows  the calculation of the MF 
surface tension. In the 
present paper we reconsider and extend Brilliantov' s work on the MF 
level of the KSSHE theory and go a step further by 
considering also  the one-loop approximation of the
theory or, more 
precisely, a simplified version of it, called the Gaussian approximation in 
the literature \cite{Ma}. In the homogeneous case it
is shown to be strictly equivalent to the random 
phase approximation (RPA) of the theory of liquids \cite{Hansen}.
The present work is hopefully a first step forwards a renormalized 
theory of the critical points of liquids at the one-loop order.

Our paper is organised as follows. In section\ \ref{KSSHE} we expose in 
details some important properties of the KSSHE transform and establish 
notably the relation between the correlation functions of the fields and that 
of the fluctuations  of densities. In section\ \ref{MF} the general MF 
equations of the  KSSHE field theory  are derived and 
their properties studied in details. As a subproduct of this formalism
we obtain exact 
bounds for the free energy and the grand potential functionals. In  
next section\ \ref{G} the Gaussian approximation is derived and shown to be 
equivalent the  RPA  theory.
In section\ \ref{LG} we reexpress the action
${\cal H}(\phi)$ under a form similar to that of the Landau-Ginzburg 
(LG) Hamiltonian of magnetic systems which allows us to map the 
liquid-gas transition on the ferromagnetic one. In this way one can 
identify clearly the order parameter of the liquid-vapour 
transition which is, at least at the MF level, nothing but the density 
of the fluid, which rules out the field-mixing hypothesis 
\cite{Rehr,Bruce}.
 In section\ \ref{CP} the MF and Gaussian 
approximations are used to study in details the critical point of the fluid. 
We conclude in section\ \ref{CONCLU}.

\section{The KSSHE Transform}
\label{KSSHE}
\subsection{The KSSHE Transform of The Boltzmann Factor}
\label{KSSHE_Weight}
We consider  the case of a simple three dimensional  fluid made of $N$
identical hard spheres of diameter $\sigma$ with additional pair interactions
denoted for convenience $-w(i,j)$. Since $w(i,j)$ is an arbitrary function 
in the core
, i.e. for $r_{ij}<\sigma$, it is always possible to introduce the following
decomposition : 

\begin{eqnarray}
\label{w}
w(i,j) & = & w_{-}(i,j)-w_{+}(i,j) \; ,  \nonumber \\
\widetilde{w}_{\pm}(k) & = & \int d^{3} \vec{r} \; \exp(i\vec{k}.\vec{r}) \; 
 w_{\pm}(r) \; >0 \;\; (\forall k)  \; .
\end{eqnarray}
In order to define safely the KSSHE transformation it is 
crucial to assume that the Fourier transforms of the pair potentials 
$w_{\pm}(r)$  do exist and are both
positive functions of $k$ (for more details see the appendix).
Moreover we demand  $\widetilde{w}_{\pm}(k=0)$ and
$w_{\pm}(r=0)$ to be both well-behaved (finite) quantities. Clearly
$w_{+}(r)$ is the repulsive part of the tail and $w_{-}(r)$ its 
attractive part. 

Obviously the decomposition\ (\ref{w}) is not unique,
with the unpleasant consequence that approximate theories can depend 
upon the pecular choice made for $w_{\pm}(r)$ inside the core. We 
must precise however what we mean by an "approximate theory". If we 
mean a systematic expansion of some physical quantity 
in powers of some small physical parameter then the "theory" should be 
independant of the decomposition\ (\ref{w}). This point was 
scrupulously considered in ref. \cite{Cai-Raim} where it was shown 
that each term of the high-temperature and low activity expansions
of the equation of state of the RPM   obtained in the 
framework of the KSSHE formalism were indeed independant of the 
regularisation of the Coulomb potential inside the core.

By contrast,
the small parameter of the loop-expansion considered in the present 
work is quite mysterious (we could have noted it $\hbar$ according to 
the tradition !) 
and certainly not related to
any physical parameter of the system \cite{Parisi,Zinn,Dowrick}. As a
consequence,  
each term of the expansion of  the free energy of the liquid
given in the following sections
depends explicitly upon the regularisation of $w(r)$ at short distances.
We shall explain later on how to deal with, and even take advantage of 
the explicit dependance of the theory upon the decomposition\ (\ref{w}) 
inside the core.

We denote by $\Omega$ the domain of volume $V$
occupied by the molecules of the fluid and by $T$ their
temperature. It is convenient to assume that $\Omega$ is a cube with
periodic boundary conditions (PBC). In a given configuration
$\omega=(N;\vec{r}_1 \ldots \vec{r}_N)$
the microscopic density of particles reads  
\begin{equation}
\widehat{\rho}(\vec{r}\vert \omega)=
\sum_{i=1}^{N} \delta^{(3)}(\vec{r}-\vec{r}_i) \; ,
\end{equation}
and the Boltzmann factor can thus be written
\begin{equation}
\label{Boltz}
\exp\left(-\beta V\left(\omega\right)\right)=
\exp\left(-\beta V_{HS}\left(\omega\right) \right)\exp(-N\nu_{S})
\exp\left(\frac{1}{2} \left< \widehat{\rho} \vert \beta w \vert\widehat{\rho} 
\right>\right)
\; ,
\end{equation}
where $\beta =1/k_BT$ ($k_B$ Boltzmann factor) 
and $\nu_S= \beta w(0)/2$ is minus the  self-energy of the
particles. From our hypothesis on $w(r)$,
$\nu_S$ is a finite quantity which depends on the regularisation of 
the potential in the core.
 In eq.\ (\ref{Boltz}) $\exp(-\beta V_{HS}(\omega))$ denotes the hard
sphere contribution to the Boltzmann factor and we have introduced the 
symbolic notation
\begin{equation}
\left< \widehat{\rho} \vert w \vert\widehat{\rho} \right> \equiv
\int_{\Omega} d^{3} \vec{r}_1 d^{3} \vec{r}_2 \; 
\widehat{\rho}(\vec{r_1}\vert \omega)
 w(\vec{r}_1, \vec{r}_2)  \widehat{\rho}(\vec{r_2}\vert \omega) \; .
\end{equation} 
With our assumptions the two quadratic forms
$ \left< \widehat{\rho} \vert w_{\pm} \vert\widehat{\rho}\right>$ are both
 positive definite and we can
 take advantage of the properties of Gaussian functional
integrals to rewrite 
\begin{eqnarray}
\label{KSSHE1}
\exp\left (+\frac{1}{2} \left< \widehat{\rho} \vert \beta  w_{-}
 \vert\widehat{\rho} \right>\right)
&=& 
\left< \exp\left(\left<\widehat{\rho}\vert \varphi_{-}\right>\right)
\right>_{\beta w_{-}}
\nonumber \\
\exp\left(-\frac{1}{2} \left< \widehat{\rho}
\vert \beta w_{+} \vert\widehat{\rho} \right>\right)
&=&
\left< \exp( i \left<\widehat{\rho}\vert \varphi_{+}\right>)\right>_{\beta w_{+}}
\; ,
\end{eqnarray}
where the scalar products 
$\left<\widehat{\rho}\vert \varphi_{\pm}\right>$  of 
the microscopic density $\widehat{\rho}$ with the two real scalar fields
$\varphi_{\pm}(\vec{r})$ are defined as  

\begin{equation}
\label{scalarproduct}
\left<\widehat{\rho}\vert \varphi_{\pm}\right>= 
\int_{\Omega} d^{3} \vec{r} \; \widehat{\rho}(\vec{r}\vert \omega)
\varphi_{\pm}(\vec{r}) \; .
\end{equation}
 The brackets $\left< \ldots  \right>_{\beta w_{\pm}}$
denote Gaussian averages over the fields $\varphi_{\pm}(\vec{r})$ i.e.
\begin{eqnarray}
\label{def}
\left< \ldots  \right>_{\beta w_{\pm}}&\equiv&
{\cal N}_{\beta w_{\pm}}^{-1} \int {\cal D} \varphi_{\pm}(\vec{r})
\ldots \exp\left(-\frac{1}{2} \left< \varphi_{\pm} \vert (\beta w_{\pm})^{-1}
\vert \varphi_{\pm} \right>\right) \; , \nonumber \\
{\cal N}_{\beta w_{\pm}} &\equiv& \int {\cal D } \varphi_{\pm}(\vec{r})
\exp\left(-\frac{1}{2} \left< \varphi_{\pm} \vert (\beta w_{\pm})^{-1}
\vert \varphi_{\pm} \right>\right) \; ,
\end{eqnarray}
where the inverse of $\beta w_{\pm}$ must be understood in the sense of
operators, i.e.
\begin{equation}
\label{inverse}
\int_{\Omega} d^{3} \vec{r}_{3}\; w_{\pm}(\vec{r}_{1},\vec{r}_{3}) 
w_{\pm}^{-1} (\vec{r}_{3},\vec{r}_{2})=\delta^{(3)} (\vec{r}_{1},\vec{r}_{2}) \: .
\end{equation}
One could object that the relations\ (\ref{KSSHE1}),\ (\ref{def}) and\
(\ref{inverse})  are rather
formal but in  simple cases - i.e. if PBC are assumed -
explicit and unambiguous
expressions of the measure
$ {\cal D } \varphi_{\pm}(\vec{r}) $ can be given. 
See \mbox{appendix A} for more details. 
In order to simplify the notations further, we introduce the
complex field $\varphi(\vec{r}) \equiv  \varphi_{-}(\vec{r}) 
+ i \varphi_{+}(\vec{r})$, define the measure 
\begin{equation}
    \label{caval}
{\cal D } \varphi(\vec{r})\equiv {\cal D } \varphi_{+}(\vec{r})
 {\cal D } \varphi_{-}(\vec{r}) \; ,
 \end{equation} 
and the Gaussian average
\begin{equation}
\left< \ldots  \right>_{\beta w} \equiv 
{\cal N }_{\beta w}^{-1} \int {\cal D } \varphi
(\vec{r}) \ldots
\exp\left(-\frac{1}{2} \sum_{\epsilon=\pm}\left< \varphi_{\epsilon}
 \vert (\beta w_{\epsilon})^{-1}
\vert \varphi_{\epsilon} \right>\right) \; ,
\end{equation}
where $ {\cal{N}}_{\beta w}\equiv {\cal{N}}_{\beta w_{+}}\times
 {\cal{N}}_{\beta w_{-}}$. We can thus rewrite the Boltzman factor\
 (\ref{Boltz}) as
 
\begin{equation}
\label{Boltz2} 
\exp(-\beta V(\omega))=\exp(-\beta V_{HS}(\omega))\exp(N\nu_{S})
\left< \exp(\left< \varphi \vert \widehat{\rho} \right>)\right>_{\beta 
w}
\; .
\end{equation}
Admittedly these notations can be misleading but they can be fully 
justified, see  the beginning of section\ \ref{LG} for a more detailed
discussion.
Eq. (\ref{Boltz2}) defines the KSSHE transform of the Botzmann factor.

\subsection{The Physical meaning of the auxilliary fields}
\label{phys_mean} 
We have seen in section\ \ref{KSSHE_Weight} 
that, in a given configuration $\omega$,
the Boltzmann factor of the fluid can be written as 
the statistical field partition function

\begin{equation}
z(\omega)\equiv \exp(\frac{1}{2} <\widehat{\rho}\vert \beta w \vert \widehat{\rho}>  )=
{\cal{N}}_{\beta w}^{-1} \int {\cal D } \varphi 
\exp(-h[\varphi]) \; ,
\end{equation}
where  $h[\varphi]$ is a Gaussian Hamiltonian  which reads
\begin{equation}
h[\varphi] = \frac{1}{2}\sum_{\epsilon=\pm}\left< \varphi_{\epsilon}
 \vert (\beta w_{\epsilon})^{-1}
\vert \varphi_{\epsilon} \right>  -\left<\varphi_{-} \vert \widehat{\rho}
\right> - i \left<\varphi_{+} \vert \widehat{\rho}
\right> \; .
\end{equation}
The saddle point of $h[\varphi]$ is defined by the set of equations
\begin{equation}
\left.\frac{\delta h}{\delta \varphi_{-}(\vec{r})}\right |_{\overline{\varphi}}
= 
\left.\frac{\delta h}{\delta \varphi_{+}(\vec{r})}\right |_{\overline{\varphi}}
=
0
\end{equation}
the solution of which is 
\begin{eqnarray}
\label{pm}
\overline{\varphi}_{+}(\vec{r}_{1}) & = & i \beta
\int_{\Omega} d^{3}\vec{r}_{2}\; w_{+}(\vec{r}_{1},\vec{r}_{2}) \; \widehat{\rho}
(\vec{r}_{2})
\nonumber \; , \\
& \equiv &  i \beta  w_{+}(\vec{r}_{1},.) * \widehat{\rho}(.) \; , \nonumber \\
\overline{\varphi}_{-}(\vec{r}_1) & = & 
 \beta   w_{-}(\vec{r}_{1},.) * \widehat{\rho}(.) \; ,
\end{eqnarray}
where the symbol $'*'$ denotes a convolution in space. Note that 
eqs\ (\ref{pm}) imply the compact formula
\begin{equation}
\overline{\varphi}(\vec{r}_{1})  =   \beta  w(\vec{r}_{1},.) * 
\widehat{\rho}(.) \; ,
\end{equation}
where $\overline{\varphi}(\vec{r})
= \overline{\varphi}_{-}(\vec{r}) + i \overline{\varphi}_{+}(\vec{r})$, or, more
explicitely
\begin{equation}
\overline{\varphi}(\vec{r})=\sum_{i=1}^{N}\beta w (\vec{r}-\vec{r}_i) \; .
\end{equation}
Therefore, at the saddle point, the KSSHE field
$\overline{\varphi}(\vec{r})$ is simply minus the
local potential energy at point $\vec{r}$. Moreover one notes  that 
\begin{equation}
h[\overline{\varphi}]=-\frac{1}{2} \left< \widehat{\rho} \vert \beta w \vert
\widehat{\rho} \right>\; .
\end{equation}
i.e. the value of $h$ at the saddle point $  h[\overline{\varphi}]$ 
coincides with  the configurational energy. Let us make the change of
variables $\varphi_{\pm}(\vec{r})=\overline{\varphi}_{\pm}(\vec{r}) + 
\delta \varphi_{\pm}(\vec{r})$, one has obviously, as a consequence of the
stationarity,
\begin{equation}
h[\varphi]=h[\overline{\varphi}] + \frac{1}{2} \sum_{\epsilon=\pm}
\left< \delta \varphi_{\epsilon}
 \vert (\beta w_{\epsilon})^{-1}
\vert \delta \varphi_{\epsilon} \right>
\end{equation}
The positivity of both operators $w^{-1}_{\pm}$ confirms 
that $( \overline{\varphi}, h[\overline{\varphi}])$
is indeed a saddle point.
For a more complicated Hamiltonian than $h[\varphi]$, the approximation
consisting in truncating the corrections to the saddle point
 value $h[\overline{\varphi]}$ at the Gaussian level is called the Gaussian
 approximation \cite{Ma,Parisi}. 
 This approximation is obviouly exact for $z(\omega)$ because
 $h[\varphi]$ is a quadratic form. Indeed a direct calculation shows 
 that  
 \begin{equation}
     \label{sp}
 z_{G} \equiv {\cal{N}}_{\beta w}^{-1} \int {\cal D }\delta \varphi(\vec{r})
\exp\left(-h[\overline{\varphi}]
-\frac{1}{2} \sum_{\epsilon=\pm}\left< \delta \varphi_{\epsilon}
 \vert (\beta w_{\epsilon})^{-1}
\vert \delta \varphi_{\epsilon} \right>\right)
=\exp\left(-h[\overline{\varphi}]\right) \equiv z(\omega) \; .
\end{equation}
Note that it follows from eq.\ (\ref{pm}) that 
$\overline{\varphi}_{+}$ is imaginary whereas we assumed the reality
of the field $\varphi_{+}$. The integrals on $\delta \varphi_{+}$ in 
eq.\ (\ref{sp}) correspond therefore to a mere translation for the contour 
of integration.
\subsection{The KSSHE Transform of The Grand Partition Function}
\label{KSSHE_GC}
Henceforward we shall 
work in the grand canonical (GC) ensemble. We denote by $\mu$ 
the chemical potential an by $\psi(\vec{r})$ an external potential. The local
chemical potential will be noted $\nu(\vec{r})=\beta(\mu-\psi(\vec{r}))$ and the
GC partition function reads :  
\begin{equation}
\label{Xi}
\Xi[\nu]=\sum_{N=0}^{\infty} \frac{1}{N!} \int_{\Omega}d1 \ldots dN \exp(-\beta
V(\omega)) \prod_{i=1}^{N}\exp\left(\nu\left(i\right)\right) \; ,
\end{equation}
where $i \equiv \vec{r}_i$ and $di \equiv d^{3}\vec{r}_i$.
$\Xi$  can be symbolically rewritten as
\begin{equation}
\label{Xi_bis}
\Xi[\nu] = \int d\mu(\omega) \exp\left(-\beta V\left(\omega\right)\right)\;
 \exp\left(\left<\nu \vert \widehat{\rho} \right> \right) \; .
\end{equation}
With these notations averages will be obtained by the formula
\begin{equation}
\left< A(\omega) \right>_{GC} \equiv 
\frac{\int d\mu(\omega)A(\omega) \exp\left(-\beta V\left(\omega\right)\right)\;
 \exp\left(\left<\nu \vert \widehat{\rho} \right> \right)}{\int d\mu(\omega) \exp\left(-\beta V\left(\omega\right)\right)\;
 \exp\left(\left<\nu \vert \widehat{\rho} \right> \right)} \; .
\end{equation}
As well known, $\Xi[\nu]$ is log-convex functional of 
$\nu$ \cite{Chayes,Percus,Caillol}. Inserting
the expression\ (\ref{Boltz2}) in eq.\ (\ref{Xi}) one obtains readily a 
result which seems to have been established  for the first time by 
Siegert \cite{Siegert}, i.e. 
\begin{equation}
\label{basic}
\Xi[\nu]= \left< \Xi_{HS}\left[\overline{\nu} \right] \right>_{\beta w} \; .
\end{equation}
where $\overline{\nu}(\vec{r})=\nu(\vec{r})-\nu_{S}+\varphi(\vec{r})$. Eq\
(\ref{basic}) tells us that the grand partition function $\Xi$ of a fluid of
decorated hard spheres is equal to the
mean value of the grand partition function $\Xi_{HS}$ of
bare hard spheres in the presence of an external stochastic field with a
Gaussian weight. 

To make some contact with statistical field theory we also introduce the
effective Hamiltonian (or action)
\begin{equation}
\label{H}
{\cal H}[\varphi]  = \frac{1}{2} \sum_{\epsilon=\pm}\left< 
\varphi_{\epsilon}
 \vert (\beta w_{\epsilon})^{-1}
\vert \varphi_{\epsilon} \right>  - 
\log \Xi_{HS}\left[\overline{\nu}\right] \; ,
\end{equation}
which allows us to write alternatively for $\Xi$
\begin{equation}
\label{Xinewlook}
\Xi[\nu]= {\cal N}_{\beta w}^{-1} \int {\cal D} \varphi(\vec{r})
 \exp(-{\cal H}[\varphi])
\: .
\end{equation}
It will be important in the sequel 
to distinguish carefully two types of statistical field
averages, 
the already defined  $<\ldots>_{\beta w}$
and the  $<\ldots>_{{\cal H}}$ that we define as
\begin{equation}
<A[\varphi]>_{{\cal H}} \equiv \frac{\int {\cal D} \varphi(\vec{r})
 \exp(-{\cal H}[\varphi])A[\varphi]}{\int {\cal D} \varphi(\vec{r})
 \; \exp(-{\cal H}[\varphi])} \; .
\end{equation}
With these definitions in mind one notes that for an arbitrary functional of
field ${\cal A}[\varphi]$ one has 
\begin{equation}
\label{means}
<A[\varphi]>_{{\cal H}}= \frac{\; \left<A[\varphi] \;  
\Xi_{HS}\left[\overline{\nu}\right] \right>_{\beta w}}{\left< \;
\Xi_{HS}\left[\overline{\nu}\right]\right>_{\beta w}} \; .
\end{equation}
\subsection{Preliminary Results on Correlation Functions.}
\label{corre}
The ordinary and connected correlation  functions of the fluid will be defined
in this paper as \cite{Hansen,Stell1,Stell2} 
\begin{eqnarray}
\label{defcorre}
G^{(n)}[\nu](1, \ldots, n) &=& \frac{1}{\Xi[\nu]}\frac{\delta^{n} \;\Xi[\nu]}
{\delta \nu(1) \ldots \delta \nu(n)}           \; ,\nonumber \\
        &=& \left< \prod_{1=1}^{n} \widehat{\rho}
	(\vec{r}_{i}  \vert \omega) \right>_{GC} \; ,\nonumber \\
G^{(n)}_{c}[\nu](1, \ldots, n) &=&  \frac{\delta^{n} \log \Xi[\nu]}
{\delta \nu(1) \ldots \delta \nu(n)} \; .
\end{eqnarray}
Our notation emphasizes the fact that the  $G^{(n)}$ 
(connected and not connected) are functionals of the local chemical potential 
 $\nu(\vec{r})$ and
functions of the coordinates $(1,\ldots, n) \equiv (\vec{r}_{1},\ldots,
\vec{r}_{n})$. We know from the theory of liquids that 
\cite{Stell1,Stell2}
\begin{equation}
G^{(n)}_c [\nu](1,\ldots,n)= G^{(n)}[\nu]( 1,\ldots,n)
- \sum \prod_{m<n}€G^{(m)}_c [\nu](i_{1},\ldots,i_{m})  \; ,
\end{equation} 
where the sum of products is carried out over all possible partitions of 
the set $(1,\ldots,n)$ into subsets of cardinal $m<n$. 
In standard textbooks\cite{Hansen} the n-body correlations are usually 
defined as functional
derivatives with respect to the activity rather than with respect to
the chemical potential. It yields to differences involving delta functions. For
instance for $n=2$ and for a homogeneous system one has 
\begin{eqnarray}
\label{G2}
G^{(2)} [\nu](1,2) &=& \rho^2 g(r_{12})+ \rho \delta(1,2) \; , \nonumber \\
G^{(2)}_c [\nu](1,2) &=& \rho^2 h(r_{12}) + \rho \delta(1,2) \; ,
\end{eqnarray}
where $\rho$ is the density of the fluid and $g(r)$ the usual pair 
distribution function at chemical potential $\nu$, finally $h=g-1$.

Of course we would like to relate the correlation functions of the microscopic
density $\widehat{\rho}$ with the correlation functions of the
KSSHE field $\varphi$  defined as 

\begin{eqnarray}
G_{\varphi}^{(n)}[\nu](1, \ldots,n) & = & \left< 
\prod_{i=1}^{n} \varphi(i)\right>_{{\cal H}} \; , \nonumber \\
G_{\varphi,c}^{(n)}[\nu](1, \ldots,n) & = &  G_{\varphi}^{n}
[\nu](1, \ldots,n)
-\sum \prod_{m<n}€G^{(m)}_{\varphi,c} [\nu](i_{1},\ldots,i_{m})  \; ,
\end{eqnarray}
From the expression\ (\ref{Xinewlook}) of $\Xi$ and the definition\
(\ref{defcorre}) of $G^{(n)}$ one infers that

\begin{equation}
 G^{(n)}[\nu](1,\ldots,n)=\Xi^{-1} \; {\cal N}_{\beta w }^{-1}
 \int {\cal D} \varphi \; \exp\left(-\frac{1}{2} \sum_{\epsilon=\pm}
 \left<  \varphi_{\epsilon}
 \vert (\beta w_{\epsilon})^{-1}
\vert \varphi_{\epsilon} \right> \right) \;  \Xi_{HS}(\overline{\nu}) \;
G^{(n)}_{HS}[\overline{\nu}](1, \ldots, n) \; ,
\end{equation} 
which yields, with the help of eq.\ (\ref{means}), to the simple relation 
\begin{equation}
\label{GGphi}
 G^{(n)}[\nu](1,\ldots,n)= \left< G^{(n)}_{HS}[\overline{\nu}]
 (1, \ldots, n)
 \right>_{{\cal H}} \; .
\end{equation}
We are only half the way since the hard sphere correlation functions
 $G^{(n)}_{HS}[\overline{\nu}]$ are
of course complicated functionals of $\varphi$ (through their dependence upon
$\overline{\nu}$). Let
us first examine the simple case $n=1$ ( $G^{(n=1)}\equiv \rho $). We first 
rewrite eq.\ (\ref{GGphi}) for $n=1$ as
\begin{eqnarray}
\label{rho}
\rho[\nu](1) &=& \left< \rho_{HS}[\overline{\nu}](1)  
\right>_{{\cal H}} \; , \nonumber \\
 &=&
 \Xi^{-1} \; {\cal N}_{\beta w}^{-1} 
 \int {\cal D} \varphi \; \exp\left(-\frac{1}{2} \sum_{\epsilon=\pm}
 \left<  \varphi_{\epsilon}
 \vert (\beta w_{\epsilon})^{-1}
\vert \varphi_{\epsilon} \right> \right) \; \frac{\delta \Xi_{HS}
(\overline{\nu})}{\delta\nu(1)} \; .
\end{eqnarray}
Then one remarks that 
\begin{equation}
\label{deriv}
\frac{\delta \Xi_{HS} (\overline{\nu})}{\delta\nu(1)}=
\frac{\delta \Xi_{HS} (\overline{\nu})}{\delta\varphi_{-}(1)}=
-i\;\frac{\delta \Xi_{HS} (\overline{\nu})}{\delta\varphi_{+}(1)} \; ,
\end{equation}
which allows us to replace the functional derivative of $\Xi_{HS}$ 
with respect to $\nu$ in the r.h.s
of eq.\ (\ref{rho}) by a derivative either with respect to $\varphi_{-}$ or 
with respect to $\varphi_{+}$. Let us do it in details in the former case.
Since for functional integrals  one has \cite{Parisi}
\begin{equation}
\int {\cal D} \varphi(\vec{r}) \frac{\delta I[\varphi]}
{\delta \varphi_{\pm}(\vec{r})} =0 \;,
\end{equation}
An integration by parts  yields the results
\begin{eqnarray}
\rho[\nu](1) &=& - \Xi^{-1} \; {\cal N}_{\beta w}^{-1} 
\int {\cal D} \varphi \; \frac{\delta 
\exp\left(-\frac{1}{2} \sum_{\epsilon=\pm}
 \left<  \varphi_{\epsilon}
 \vert (\beta w_{\epsilon})^{-1}
\vert \varphi_{\epsilon} \right> \right)}{\delta \varphi_{-}(1)}
\; \Xi_{HS}(\overline{\nu}) \; ,
 \nonumber \\
 &=&
  (\beta w_{-})^{-1}(1,.) * \left<  \varphi_{-}(.) \right>_{{\cal H}} \; .
\end{eqnarray}
Repeating the above derivation with
the field $\varphi_{+}$ instead of $\varphi_{-}$
and making use of the compact
notations previously defined one finally arrives at the set of relations 
\begin{mathletters}
\begin{equation}
\label{aa}
\left<  \varphi_{-}(1) \right>_{{\cal H}}  =  \beta w_{-}(1,.) *
\left<  \rho_{HS}[\overline{\nu}](.) \right>_{{\cal H}} \; , 
\end{equation}
\begin{equation}
\label{bb}
 \left<  \varphi_{+}(1) \right>_{{\cal H}}  =   i \beta w_{+}(1,.) *
 \left<  \rho_{HS}[\overline{\nu}](.) \right>_{{\cal H}} \; ,
 \end{equation}
\begin{equation}
\label{cc}
 \left<  \varphi(1) \right>_{{\cal H}} =   \beta w(1,.)*
 \left<  \rho_{HS}[\overline{\nu}](.) \right>_{{\cal H}} \; .
\end{equation}  
\end{mathletters} 
For a homogeneous system one therefore has
$\left<  \varphi \right>_{{\cal H}}=\beta \widetilde{w}(0) \rho $. 
It can be noticed
that  eq.\ (\ref{cc}) is analogous to a relation derived by Callen for an 
array of spins $S_{i}$  \cite{Callen}, namely
$<S_i>= <\tanh(\beta h_i + \beta J_{ij} S_j)>$ ($h_i$ magnetic field on
the site $i$, $J_{ik}$ coupling constant between sites $i \text{ and } j$). 
Callen's exact relation can be used to obtain the MF equations of spin systems
by neglecting all correlations, which yields the well-known MF equations 
$\overline{S}_{i}=\tanh(\beta h_i + \beta J_{ij} \overline{S}_j)$ where
$\overline{S}_i$ is the MF magnetisation of site $i$. Proceeding in an analogous
way with eq\ (\ref{cc}) on obtains the MF equations for the
KSSHE field $\varphi$:
\begin{equation}
\label{MF1}
\overline{\varphi}(1) = \beta w(1,.) * \rho_{HS}
[\nu -\nu_{S} + \overline{\varphi}]( . ) \; ,
\end{equation} 
which will be derived more rigorously in next section. 

This method of functional integration by parts can be used {\em a priori} 
at all orders so that to derive
relations between the correlations $G^{(n)}_{\varphi}$
of the field $\varphi$ and  the correlations $G^{(n)}$ of the microscopic
density $\widehat{\rho}$.  We just quote the result for $n=2$.
\begin{eqnarray}
\label{GGphi2}
G^{(2)}_{\varphi}[\nu](1,2) & =& \beta w(1,2)
 + \beta w(1,.) * G^{(2)}[\nu](.,.) * \beta w(.,2) \; , \nonumber \\
 G^{(2)}_{\varphi,c}[\nu](1,2) & =& \beta w(1,2)
 + \beta w(1,.) * G^{(2)}_{c}[\nu](.,.) * \beta w(.,2) \; .
\end{eqnarray}
We leave the derivation of eqs.\ (\ref{GGphi2}) by the method of 
integration by parts as an exercise for the reader. The result will 
be rederived and generalised to the case $n\geq 3$ 
by a more elegant method in section\ \ref{LG}.
\section{Mean Field Theory}
\label{MF}
\subsection{The grand-canonical free energy}
\label{Kohn-Sham}
We define the MF theory or saddle point approximation by the equation
\begin{equation}
\Xi_{MF}(\nu)\equiv \exp(-{\cal H}(\overline{\varphi})) \; ,
\end{equation}
where at $\varphi=\overline{\varphi}$ the action ${\cal H}$ is minimum. 
It there
are several local minima then one retains the absolute minimum.
The stationary condition reads
\begin{equation}
    \label{statio}
\left.\frac{\delta {\cal H}}{\delta \varphi_{\epsilon}(\vec{r})}
\right|_{\overline{\varphi}}=0 \; \; \; (\epsilon = \pm) \; \; .
\end{equation}
It follows readily from the expression \ (\ref{H}) of ${\cal H}$ 
that these eqs.  may be written  
\begin{eqnarray}
\label{eqCM}
\overline{\varphi}_{-}(1) & = & \beta w_{-}(1,.)*\rho_{HS}
[\overline{\nu}](.) \;,
\nonumber \\
\overline{\varphi}_{+}(1) & = & i\beta w_{+}(1,.)*
\rho_{HS}[\overline{\nu}](.) \;.
\end{eqnarray}
By combining linearly  eqs.\ (\ref{eqCM}) one  
recovers the MF eq.\ (\ref{MF1}) for $\varphi$. The similarity between the MF 
eqs.\ (\ref{eqCM}) and (\ref{MF1}) in one hand,
and that which give the saddle point of $h[\phi] $ in 
a given configuration of the phase space (cf. eqs.\ (\ref{pm})) on the 
other hand, is striking although expected. 

Making use of eqs\ (\ref{eqCM}) it
is then easy to show that
\begin{equation}
\log \Xi_{MF}[\nu]=\log \Xi_{HS}[\overline{\nu}] 
-\frac{1}{2} 
\left< \rho_{HS}[\overline{\nu}] \vert \beta w \vert \rho_{HS}[\overline{\nu}]
\right> \; .
\end{equation}
The MF density at point $\vec{r}_1$ is obtained by taking the functional
derivative of $\log \Xi_{MF}[\nu]$ with respect to $\nu(1)$. It yields
\begin{eqnarray}
\label{roMF}
\rho_{MF}[\nu](1) & = & \frac{\delta \log \Xi_{MF}[\nu]}{\delta \nu(1)}
\;  \nonumber \\ 
& = & \rho_{HS}[\overline{\nu}](1) +
\left<\rho_{HS}[\overline{\nu}] \; \vert  \; 
\frac{\delta \overline{\varphi}}{\delta \nu(1)} -\beta w * \frac
{\rho_{HS}}{\delta \nu(1)} \right> \; .
\end{eqnarray}
The scalar product in the r.h.s of eq.\ (\ref{roMF}) 
vanishes as a consequence of the
stationarity conditions\ (\ref{statio}) and, finally, one finds
\begin{equation}
\label{roMF2}    
\rho_{MF}[\nu](1) = \rho_{HS}[\overline{\nu}](1)
\end{equation}
We are now in position to compute the MF 
grand-canonical (MFGC) free energy as the Legendre Transform of
 $\log \Xi_{MF}[\nu]$ with respect to the local chemical potential 
 $\nu(\vec{r})$
\begin{equation}
\label{freeener}
\beta {\cal A}_{MF}[\rho_{MF}] = \left<\rho_{MF} \vert \nu \right>
- \log \Xi_{MF}[\nu]
\end{equation}
 One finds after simple algebra that ${\cal A}[\rho]$,
 as a functional of $\rho$, can
 be expressed as 
\begin{equation}
 \label{freeener2} 
 \beta {\cal A}_{MF}[\rho]=\beta {\cal A}_{HS}[\rho] + \int_{\Omega}
 d^{3}\vec{r} \;
 \nu_{S} \rho(\vec{r}) -\frac{1}{2} \left< \rho \vert \beta w \vert \rho
  \right> \: ,
\end{equation}
where ${\cal A}_{HS}[\rho]$ is the exact GC free energy functional 
of the hard
spheres at the density $\rho$. For a homogeneous system 
${\cal A}_{MF}(\rho)$
is merely the  MF  free energy of the system in the GC
ensemble.
\subsection{The Mean Field Density Functionals}
We recall first some important properties of the functionals $\log \Xi [\nu]$
and ${\cal A}[\rho]$. Under quite general conditions it can be shown that
the logarithm of the exact 
grand-partition function $\log \Xi [\nu]$ is a convex functional of 
the local chemical  potential $\nu(\vec{r})$ and that the exact
GC free energy 
${\cal A}[\rho]$ is  a convex functional of the density
 $\rho(\vec{r})$\cite{Chayes,Percus,Caillol}. It must be stressed 
 that, for a finite system, ${\cal A}[\rho]$ differs from the {\em 
 canonical} free energy ${\cal A}_{C}€[\rho]$ which is not a convex 
 functional of $\rho(\vec{r})$ notably in the two-phases region due to 
 interfacial effects. 
Moreover
 $\log \Xi [\nu]$ and  ${\cal A}[\rho]$   constitute a pair of Legendre 
transforms, an important  property which can be expressed as 
\begin{eqnarray}
 \label{4} 
 \beta {\cal A}[\rho] & =& \sup_{\nu \in {\cal U}} 
 \left( \langle \rho  \vert \nu \rangle - \log \Xi [\nu] \right) \; \; (\forall  
 \rho \in {\cal R}) \; ,  \\
 \label{5} 
 \log \Xi [\nu] &=& \sup_{\rho \in {\cal R}} 
 \left( \langle \rho  \vert \nu \rangle -  \beta {\cal A}[\rho] \right)
\; \; (\forall    \nu \in {\cal U}) \; .
\end{eqnarray}
It can also be shown that 
the sets of physical densities ${\cal R}$ and that of local chemical potentials
 ${\cal U}$ are both convex sets \cite{Chayes,Caillol}. 
 The so-called Young inequalities
  follows directly from eqs\ (\ref{4}),\ (\ref{5}) and read 
 \begin{equation}
\label{Young}  
\beta {\cal A}[\rho] + \log \Xi [\nu] \geq \langle \rho 
\vert \nu \rangle \; \; (\forall  
 \rho \in {\cal R}, \forall    \nu \in {\cal U} ) \;   \\ .
\end{equation} 
Finally we recall that a necessary and sufficient condition
 for the convexity of $\log \Xi [\nu]$
and ${\cal A}[\rho]$  is that their second order functional derivatives are
positive operators, i.e. \cite{Chayes,Percus,Caillol}
\begin{eqnarray}
\label{GetC}
\left< \delta \nu \vert \frac{\delta^{(2)} \log \Xi[\nu]}{\delta \nu(1)
\delta \nu(2)} \;(\equiv G^{(2)}_c )[\nu] \vert \delta \nu \right> 
& \geq &  0 \; \;(\forall \nu, \delta \nu \in {\cal U}) \;, \nonumber \\
\left< \delta \rho \vert \frac{\delta^{(2)} \beta {\cal A}[\rho] }
{\delta \rho(1)
\delta \rho(2)} \; (\equiv -\widehat{C}^{(2)}[\rho] ) \vert
 \delta \rho \right> 
& \geq &  0 \; \; (\forall \rho, \delta \rho \in {\cal R})\;.
\end{eqnarray} 
Moreover the two-body direct correlation function 
$\widehat{C}^{(2)}[\rho] $
 is minus the inverse
of the connected two-body correlation function $G^{(2)}_c[\nu] $ i.e.
\begin{equation}
\label{invGC}
G^{(2)}_c[\nu] (1,.)*\widehat{C}^{(2)}[\rho] (.,2)=-\delta(1,2) \;
\end{equation}
where it must be stressed that $\nu[\rho]$ is the (unique) 
local chemical potential corresponding to the density profile $\rho$.
We have now at our disposal 
all the mathematical tools necessary to prove the two following theorems
\begin{itemize}
\item if $w_{+}=0$ (attractive case) $\beta {\cal A}_{MF}[\rho]$ is an
upper bound of $\beta {\cal A}[\rho]$
\item if $w_{-}=0$ (repulsive case) $\beta {\cal A}_{MF}[\rho]$ is a 
lower bound of $\beta {\cal A}[\rho]$
\end{itemize}
Let us consider first the attractive case ($w_{+}=0$).
We start with the fundamental relation\ (\ref{Xi}) which reads in this case
\begin{equation}
\Xi[\nu]=\left< \exp \left(\log \Xi_{HS} [\nu -\nu_{S} + 
\varphi] \right) \right>_{\beta w_{-}} \; \; (\forall \nu \in {\cal U}) \; .
\end{equation} 
We now apply Young's inequalities\ (\ref{Young})
to $\log \Xi_{HS}$ which yields
\begin{eqnarray}
\Xi[\nu] & \geq & \left< \exp \left( -\beta {\cal A}[\rho] +
\left< \rho \vert \nu -\nu_{S} + \varphi_{-} \right> 
\right) \right>_{\beta w_{-}} \; \;
 (\forall \nu \in {\cal U}, \forall \rho \in {\cal R}) \nonumber \\
 &\equiv&\exp \left( -\beta {\cal A}_{HS}[\rho]+ \left< \rho \vert \nu -\nu_{S}
 \right> +\frac{1}{2}  
 \left< \rho \vert \beta w_{-} \vert \rho \right>
 \right) \; \; (\forall \nu \in {\cal U}, \forall \rho \in {\cal R}) \; .
 \end{eqnarray}
where we have made use of the property\ (\ref{KSSHE1}) of Gaussian 
integrals. Taking the
logarithm one arrives at
\begin{equation}
\beta {\cal A}_{MF}[\rho] \equiv
\beta {\cal A}_{HS}[\rho] + \left< \rho \vert \nu_{S} \right> -
\frac{1}{2}  
 \left< \rho \vert \beta w_{-} \vert \rho \right> \geq 
 \left< \rho \vert \nu \right>-\log \Xi[\nu] ; \;
  ; \;
  (\forall \nu \in {\cal U}, \forall \rho \in {\cal R}) \; .
\end{equation}
which implies
\begin{equation}
    \label{th1}
\beta {\cal A}_{MF}[\rho]\geq \sup_{\nu \in {\cal U}} 
 \left( \langle \rho  \vert \nu \rangle - \log \Xi [\nu] \right)
 \equiv \beta {\cal A}[\rho] 
 \; \; ( w_{+}\equiv 0, \;\forall \rho \in {\cal R}) \; .
\end{equation} 
Good approximate functionals $\beta {\cal A}_{HS}[\rho]$ 
are available in the
literature for the hard sphere fluid. Our  $\beta {\cal A}_{MF}[\rho]$ could 
be of some use to deal with  systems of decorated hard spheres. Note that 
$\beta {\cal A}_{MF}[\rho]$ is also a functional of the pair potential $\beta
w_{-}(r)$ in the core ($r\leq \sigma)$; therefore an optimized version
of the MFGC free energy can be obtained by minimizing
$\beta {\cal A}_{MF}[\rho]$
with respect to $w_{-}(r)$ in the core.

The repulsive case  ($w_{-}=0$) is more tricky. One first remarks 
that, in this case, one has
\begin{equation}
\Xi[\nu]=\left< \exp \left(\log \Xi_{HS} [\nu -\nu_{S} + i 
\varphi_{+}] \right) \right>_{\beta w_{+}} \; \; (\forall \nu \in {\cal U}) \; .
\end{equation}
which implies
\begin{equation}
\Xi_{HS}[\nu]=\left< \exp \left(\log \Xi [\nu + \nu_{S} + 
\varphi_{+}] \right) \right>_{\beta w_{+}} \; \; (\forall \nu \in {\cal U}) \; .
\end{equation}
We are thus led back to the previous case.
We therefore apply  now Young's inequalities\ (\ref{Young})
to $\log \Xi$ which yields
 \begin{equation}
\Xi_{HS}[\nu] \geq \exp\left(-\beta {\cal A}[\rho] +
\left< \rho \vert \nu + \nu_{S} \right> +
\frac{1}{2}
\left< \rho \vert \beta w_{+} \vert \rho \right> \right) 
\; \;
  (\forall \nu \in {\cal U}, \forall \rho \in {\cal R}) \; , 
\end{equation}
where we made use, once again of the fundamental property of Gaussian 
integrals (cf eq.\ (\ref{KSSHE1})).
Taking the logarithm we are thus led to the following inequalities
\begin{equation}
\label{uiui}
\beta {\cal A}[\rho] \geq -\log \Xi_{HS}[\nu] +
\left< \rho \vert \nu + \nu_{S} \right>
-\frac{1}{2}
\left< \rho \vert \beta w \vert \rho \right>   \;
  (\forall \nu \in {\cal U}, \forall \rho \in {\cal R}) \; ,
\end{equation} 
where we have noted that, in the repulsive case, $w_{+}=-w$. 
Since the inequality\ (\ref{uiui}) is valid for any $\nu \in {\cal U}$ for a
given $\rho \in {\cal R}$ it is also true for $\nu^{*}=
\sup{\nu} \; ( \nu \in {\cal U})$ which
yields the desired result
\begin{equation}
    \label{th2}
\beta {\cal A}[\rho] \geq \beta {\cal A}_{MF}[\rho] \; \;
 ( w_{-}\equiv 0, \; \; \forall \rho \in {\cal R}) \; . 
\end{equation}
In the repulsive case  an optimized version of the MFGC
free energy can thus be obtained by maximizing 
$\beta {\cal A}_{MF}[\rho]$
with respect to $w_{+}(r)$ in the core. 

In the general case, i.e. when $w=w_{-}-w_{+}$ ($w_{-}\neq 0$ and 
$w_{-}\neq 0$)  ${\cal A}_{MF}[\rho] $ is neither an upper or a lower 
bound for the exact GC free energy. Moreover, in the homogeneous case, 
there is no clear relation between ${\cal A}_{MF}[\rho] $ and the
Gibbs-Bogoliubov bounds\cite{Hansen} for the free energy.
\subsection{The Mean Field Correlation Functions}
\subsubsection{The connected two-points correlation function}
The MF connected two-points correlation function will be 
defined according to eq.\ (\ref{defcorre}), i.e. as
\begin{eqnarray}
\label{GMF}
G^{(2)}_{MF,c}[\nu](1,2) & =&  \frac{\delta^{2} \log \Xi_{MF}[\nu]}
{\delta \nu(1) \; \delta \nu(2)} \; , \nonumber \\
&=&\frac{\delta \rho_{MF}[\nu](2)} 
{ \delta \nu(1)} \; .
\end{eqnarray}
Since $\rho_{MF}[\nu](2)=\rho_{HS}[\nu-\nu_{S}+\overline{\varphi}](2)$
(cf. eq.\ (\ref{roMF2}))
the MF density $\rho_{MF}[\nu]$ is a functional of the 
local chemical potential $\nu$ directly and also through the KSSHE fied
$\overline{\varphi}[\nu]$. Therefore one has
\begin{equation}
G^{(2)}_{MF,c}[\nu](1,2)  =G^{(2)}_{HS,c}[\overline{\nu}](1,2) +
\int d3 \; G^{(2)}_{HS,c}[\overline{\nu}](1,3)
\frac{\delta \overline{\varphi}(3)}
{\delta \nu(2)} \; .
\end{equation} 
We note that it follows from from the MF eq.\ (\ref{MF1}) that
\begin{equation}
 \frac{\delta \overline{\varphi}(3)}{\delta \nu(2)}
 =
 \beta w (3,.) * G^{(2)}_{MF,c}[\nu](.,2)
\end{equation}     
from which we infer the relation
\begin{equation}
G^{(2)}_{MF,c}[\nu] = G^{(2)}_{HS,c}[\overline{\nu}] + 
G^{(2)}_{HS,c}[\overline{\nu}]*\beta w *G^{(2)}_{MF,c}[\nu] 
\end{equation} 
which can be solved formally to give 
\begin{equation} 
  \label{Grpa1}   
G^{(2)}_{MF,c}[\nu] = (1-\beta w * G^{(2)}_{HS,c}[\overline{\nu}] )^{-1}
*G^{(2)}_{HS,c}[\overline{\nu}] \; . 
\end{equation} 
For a homogeneous system the Fourier transform of $G^{(2)}_{MF,c}$ 
therefore reads as 
\begin{equation}
    \label{Grpa2}
\tilde{G}^{(2)}_{MF,c}[\nu](k)= \frac{\tilde{G}^{(2)}_{HS,c}
[\overline{\nu} ](k)}
{1- \beta \tilde{w}(k)\tilde{G}^{(2)}_{HS,c}
[\overline{\nu} ](k) } \; .
\end{equation}
\subsubsection{The direct two-points correlation function}
The  MF two-points direct correlation function will be 
defined according to eq.\ (\ref{GetC}), i.e. as
\begin{equation}
\label{CMF}
\widehat{C}^{(2)}_{MF}[\rho](1,2)=-\frac{\delta^{2}
\beta {\cal A}_{MF}[\rho] }{\delta \rho(1)
\delta \rho(2)} \; .
\end{equation}
In the terminology of statistical field theory
$\widehat{C}^{(2)}_{MF}[\rho](1,2)$ is a vertex function, it
is related to the usual 
direct correlation function $c(1,2)$ by the relation 
$\widehat{C}^{(2)}_{MF}[\rho](1,2)= 
c_{MF}[\rho](1,2) -\delta(1,2)/\rho(1)$ \cite{Stell1,Stell2}.
It readily follows from the expression\ (\ref{freeener2}) of the MFGC
 free energy that we have
\begin{equation}
\label{Crpa}
\widehat{C}^{(2)}_{MF}[\rho](1,2)=\widehat{C}^{(2)}_{HS}[\rho](1,2)
+ \beta 
w(1,2) \; .
\end{equation}
Some remarks are at order. Firstly, it can be checked that the  
mean-field operators $-\widehat{C}^{(2)}_{MF}$ is indeed the inverse
of $G^{(2)}_{MF,c}$; more precisely one checks readily that, formally
\begin{equation}
    \label{OZ}
\widehat{C}^{(2)}_{MF}[\rho](1,.) *G^{(2)}_{MF,c}[\nu](.,2)=
-\delta(1,2) \; ,
\end{equation}
where it must be stressed that in the above equation $\rho$ denotes the 
mean-field density at the local chemical potential $\nu$, i.e. $\rho \equiv 
\rho_{HS}[\overline{\nu}]$. For a homogeneous system the 
Ornstein-Zernike (OZ) eq.\ (\ref{OZ}) 
can be written simply in Fourier space as $\tilde{G}^{(2)}_{MF,c}[\nu](k)  
\tilde{\widehat{C}}^{(2)}_{MF,c}[\rho](k)=-1$. Secondly it must 
be stressed that the expression\ (\ref{Grpa1}) of the connected 
correlation function $G^{(2)}_{MF,c}$ is rather formal. Its validity, 
and thus the validity of the OZ relation\ (\ref{OZ}) as well, is guaranteed only 
is the inverse of the operator $1-\beta w * G^{(2)}_{HS,c}[\nu]$ does 
exist. It can be expected to be true only if $T > T_{c}$ where $T_{c}$ is 
the (mean field) critical temperature. Below $T_{c}$ the first 
derivatives of the MF grand 
potential with respect to $\nu$ have discontinuities at the chemical 
potential corresponding to  the coexistence of the vapour and liquid 
phases and the second derivatives 
(i.e. $ G^{(2)}_{MF,c}[\nu]$) are not defined.
\subsubsection{Rigorous bounds for $Log \;\Xi[\nu]$} 
It transpires from the above discussion that for $T>T_{c}$ 
the MF correlation 
function $ G^{(2)}_{MF,c}[\nu]$ is a positive operator, i.e. that
\begin{equation}
\label{Gpos}
(\forall \nu, \forall h \in {\cal U}) \; \; 
\left < h \vert G^{(2)}_{MF,c}[\nu] \vert h \right > \geq 0 \; .
\end{equation}
It follows then from the OZ eq.\ (\ref{OZ}) that for $T > T_{c}$ and 
for all densities $\rho \in {\cal R} $, $\widehat{C}^{(2)}_{MF,c}[\rho]$ is also a positive 
operator. Consequently, both $\log \Xi_{MF}[\nu]$ and $\beta {\cal 
A}_{MF}[\rho]$ are convex functional of respectively the generalised 
chemical potential $\nu$ and the density $\rho$. Moreover they 
constitute a pair of Legendre transforms by construction
and thus satisfy to the eqs.\ 
(\ref{4}),\ (\ref{5}). Exact bounds for $\log \Xi[\nu]$, valid for 
$T>T_{c}$,  can be deduced from this property

Let us start with the case of a repulsive pair potential (i.e. 
$w_{-}\equiv 0$). In this case $\beta {\cal A}[\rho] \geq 
\beta {\cal A}_{MF}[\rho]$ for all $\rho$ as shown by eq.\ (\ref{th2}).
Therefore, for $T>T_{c}$ 
\begin{eqnarray}
\log \Xi_{MF} [\nu] &=& \sup_{\rho \in {\cal R}} 
 \left( \langle \rho  \vert \nu \rangle -  \beta {\cal A}_{MF}[\rho] \right)
\; \; (\forall    \nu \in {\cal U}) \; , \nonumber \\
&\geq & \sup_{\rho \in {\cal R}} 
 \left( \langle \rho  \vert \nu \rangle -  \beta {\cal A}[\rho] \right)
\; \; (\forall    \nu \in {\cal U}) \; ,
\end{eqnarray}
from which we conclude that 
\begin{equation}
\label{th3} 
\log \Xi_{MF}[\nu] \geq \log \Xi[\nu] \; \;
(T>T_{c},\; w_{-}\equiv 0,  \; \forall \nu 
\in {\cal U}) \; . 
\end{equation}
In the case of a purely attractive pair potential (i.e. $w_{+}\equiv 
0$) it can be shown similarly that 
\begin{equation}
\label{th4} 
\log \Xi_{MF}[\nu] \leq \log \Xi[\nu] \; \;
(T>T_{c},\; w_{+}\equiv 0,  \; \forall \nu 
\in {\cal U}) \; . 
\end{equation}
\section{The Gaussian Approximation}
\label{G}
\subsection{The grand potential}
Let us define the Gaussian approximation of the KSSHE field theory 
in the following way \cite{Ma}. First we write 
\begin{eqnarray}
 \varphi_{-}&=& \overline{\varphi}_{-} + \xi_{-} \; , \nonumber \\   
  \varphi_{+}&=& \overline{\varphi}_{+} + \xi_{+} \; ,
\end{eqnarray}
where $\xi_{-}$ and  $\xi_{+}$ are real scalar fields  ($ \xi $ will be 
defined as $\xi = \xi_{-} +i  \xi_{+} $) and then we expand functionally
the action ${\cal H}[\varphi]$ (cf eq.\ (\ref{H}) ) up to second order 
in $\xi_{\pm}$ around the MF solution.
In this way the exact action ${\cal H}[\varphi]$ is 
replaced by an approximate action ${\cal H}_{G}[\varphi]$ defined 
symbolically as
\begin{equation}
    \label{HG}
  {\cal H}_{G}[\varphi] = {\cal H}[\overline{\varphi}] +
  \frac{1}{2} \int_{\Omega}d(1)\; d(2) \;
  \left. \frac{\delta^{2} \; {\cal H} }{\delta \varphi (1) \delta \varphi 
  (2) } \right \vert_{\overline{\varphi}} \; \xi(1) \xi(2)  \; .
\end{equation}
Note that the terms linear in $\xi$ are absent from eq.\ (\ref{HG}) as a 
consequence of the stationarity condition\ (\ref{statio}). Taking 
into account the explicit expression\ (\ref{H}) of ${\cal H}$ we 
have more precisely
\begin{equation}
    \label{HG2}
{\cal H}_{G}[\varphi] = {\cal H}[\overline{\varphi}] +
\frac{1}{2}\sum_{\epsilon=\pm}€ \left< \xi_{\epsilon}
\vert \Delta_{\epsilon}^{-1} \vert \xi_{\epsilon} \right >
- i \left< \xi_{+} \vert G_{HS,c}^{(2)}[\overline{\nu}]
\vert \xi_{-} \right > \; ,
\end{equation}
where the operators $\Delta_{\pm}$
are functionals of $\overline{\nu}=\nu -\nu_{S} + \overline{\varphi}$ 
and are defined as
\begin{equation}
    \Delta_{\pm}^{-1} \equiv (\beta w_{\pm})^{-1}
    \pm  G_{HS,c}^{(2)}[\overline{\nu}]
    \; .
\end{equation}
or equivalently
\begin{equation}
    \Delta_{\pm}= 
    \beta w_{\pm}*(1  \pm  G_{HS,c}^{(2)}[\overline{\nu}]
    *\beta w_{\pm})^{-1} 
    \; .
\end{equation}
Clearly the Gaussian approximation 
makes sense only if both operators $\Delta_{\pm}$ 
are positive, which will be assumed henceforth. 

The Gaussian grand partition function  is obtained by replacing 
${\cal H}[\varphi]$ by its Gaussian approximation 
${\cal H}_{G}[\varphi]$ in the expression\ (\ref{Xinewlook}) 
of $\Xi[\nu]$, which gives
\begin{equation}
    \Xi_{G}[\nu] =
    \exp \left(  -{\cal H}[\overline{\varphi}]\right)  
    {\cal N}_{\beta w}^{-1} \int {\cal D}\xi_{+}\; {\cal D}\xi_{+}
    \exp \left( 
  - \frac{1}{2} \sum_{\epsilon=\pm}€ \left< \xi_{\epsilon}
  \vert \Delta_{\epsilon}^{-1} \vert \xi_{\epsilon} \right >
+ i \left< \xi_{+} \vert G_{HS,c}^{(2)}[\overline{\nu}]
\vert \xi_{-} \right >  
    \right)
\end{equation}    
$\Xi_{G}[\nu] $ can be explicitly computed in the case of 
a   homogeneous system.
For instance, by first 
performing the Gaussian integrals on the variable $\xi_{+}$ 
then on $\xi_{-}$, one easily obtains that
\begin{equation}
    \label{X1}
    \Xi_{G}[\nu] = \exp \left(  -{\cal H}[\overline{\varphi}]\right) 
    {\cal N}_{\beta w}^{-1} \; {\cal N}_{\Delta_{+}} \; 
    {\cal N}_{X_{-}} \; ,
\end{equation}    
or, by reverting the order of integrations on $\xi_{+}$ and $\xi_{-}$ , that
\begin{equation}
       \label{X2}
    \Xi_{G}[\nu] = \exp \left(  -{\cal H}[\overline{\varphi}]\right) 
    {\cal N}_{\beta w}^{-1} \; {\cal N}_{\Delta_{-}} \; 
    {\cal N}_{X_{+} }\; .
\end{equation} 
The operators $X_{\pm}$ which enters eqs.\ (\ref{X1},\ref{X2}) 
being defined by the relations
\begin{equation}
    X_{\pm}^{-1} =
    \Delta_{\pm}^{-1}  + G_{HS,c}^{(2)}[\overline{\nu}]*
    \Delta_{\mp}*G_{HS,c}^{(2)}[\overline{\nu}] \; .
\end{equation} 
Since ${\cal N}_{\beta w}={\cal N}_{\beta w_{+}}
{\cal N}_{\beta w_{+}}$ we see that only ratios of the normalization 
constants ${\cal N}$ of  various operators enter the 
expressions\ (\ref{X1}, \ref{X1}) of $\Xi_{G}[\nu] $. These ratios are 
easily evaluated with the help the eq.\ (\ref{norma}) of appendix A .
For $\log \Xi 
\equiv V \beta P$ holds for a homogeneous system,
one obtains finally for the 
pressure $P_{G}$ 
\begin{eqnarray}
       \label{PG}
       \beta P_{G}(\nu) &=&  \beta P_{MF}(\nu)  +
    \frac{1}{2}   \int \frac{d^3 \vec{k}}{(2 \pi)^3} \log \left( 
       \frac{\widetilde{\Delta}_{-}(k) \widetilde{X}_{+}(k) }
       {\beta \widetilde{w}_{+} (k)  \beta \widetilde{w}_{-} (k) }\right)
       \; \nonumber \\
        &=& \beta P_{HS}(\overline{\nu}) -\frac{1}{2} \beta 
        \widetilde{w}(0) \rho_{HS}^{2}(\overline{\nu}) +
 \frac{1}{2} \int \frac{d^3 \vec{k}}{(2 \pi)^3} \log \left( 
 1- \widetilde{w}(k)  \widetilde{G}^{(2)}_{HS,c}[\overline{\nu}](k)\right)
 \; .
\end{eqnarray}	
\subsection{The correlation functions and the free energy}
The mean values of the type\ (\ref{means}) with the full action
${\cal H}$ replaced by its Gaussian approximation ${\cal H}_{G}$ will 
be denoted by $<\ldots >_{G}$. Clearly one has $<\xi_{\pm}>_{G}=0$ 
from which we infer  that 
\begin{equation}
   \rho_{G}[\nu] = \rho_{MF}[\nu] = \rho_{HS}[\overline{\nu}] \; .
\end{equation}   
The various pair correlations of 
the KSSHE fields $\varphi_{\pm}$ are obtained by applying Wick's 
theorem (cf eq.\ (\ref{Wick})) with the result
\begin{mathletters}
    \label{uu}
    \begin{eqnarray}
	\left<  \varphi_{\pm}(1) \varphi_{\pm}(2)\right>_{G,c}€&=&
	X_{\pm}(1,2) \; , \\
                      \left<  \varphi_{+}(1) \varphi_{-}(2)\right>_{G,c}€&=&
 i \Delta_{+}(1,.)*G^{(2)}_{HS,c}[\overline{\nu}](.,.)*   
\left<  \varphi_{-}(.) \varphi_{-}(2)\right>_{G,c}€  \; .
    \end{eqnarray}	   
 \end{mathletters}   
After some algebraic manipulations more easily effectuated in 
Fourier space, one obtains the more transparent relations 
\begin{mathletters}
    \label{uu9}
    \begin{eqnarray}
    \left<  \varphi_{\pm}(1) \varphi_{\pm}(2)\right>_{G,c}€&=& \beta 
    w_{\pm}(1,2)  \mp  \beta w_{\pm}(1,.)* G^{(2)}_{MF,c}[\nu](.,.)*
    \beta w_{\pm}(.,2) \; , \\
 \left<  \varphi_{+}(1) \varphi_{-}(2)\right>_{G,c}   &=& i 
 \beta w_{+}(1,.)* G^{(2)}_{MF,c}[\nu](.,.)*
    \beta w_{-}(.,2) \; ,  \\
 \left<  \varphi(1) \varphi (2)\right>_{G,c}€&=& \beta 
    w(1,2)  +  \beta w(1,.)* G^{(2)}_{MF,c}[\nu](.,.)*
    \beta w(.,2) \; .   
    \end{eqnarray}	   
 \end{mathletters}  
 The comparison of these formula with eqs.\ (\ref{GGphi2}) shows that 
 $G^{(2)}_{MF}\equiv G^{(2)}_{G}$ a well known property 
 of the Gaussian approximation\cite{Ma}. Recall however that we have defined 
 $G^{(2)}_{MF}$ as the second functional derivative of $\Xi_{MF}[\nu]$
 with respect to $\nu$. The second functional derivative of $\Xi_{G}[\nu]$
with respect to $\nu$ differs of $G^{(2)}_{G}$ as defined above by 
terms involving the three-body correlation function of the HS reference 
fluid. This difference between the two possible definitions of 
$G^{(2)}_{G}$  coincides in fact with the so-called one-loop 
correction to $G^{(2)}_{G}$ which play an important role in the 
renormalized theory of the critical point (such a theory remains 
however to be worked out explicitly !). The previous remarks could 
be extented to all correlation functions; for instance at the 
one-loop order the density does not coincides with the MF or Gaussian
result. 

Since $\rho_{G}=\rho_{MF}$ the Legendre transform of $\log \Xi_{G}[\nu]$, 
i.e. the free energy of a homogeneous system 
in the Gaussian approximation is easily derived. 
Let $f_{G}\equiv {\cal A}_{G}/V$ be the specific free energy; we have
\begin{mathletters}
    \label{freeG}
    \begin{eqnarray}
    \beta f_{G}(\rho) &=& \beta f_{MF}(\rho)  +\beta \Delta f(\rho)    \; , \\
   \beta f_{MF}(\rho) &=& \beta f_{HS}(\rho) - \frac{\beta}{2}  
   \widetilde{w}(0) \rho^{2} +\frac{\beta}{2} \rho w(0)  \; , \\
   \beta \Delta f(\rho) &=& 
   \frac{1}{2} \int \frac{d^{3}\vec k}{(2 \pi)^3} \;
   \log \left(1-\beta \widetilde{w}(k)\widetilde{G}^{(2)}_{HS,c}[\rho]( 
   k )\right) \; . \label{freeGc}
    \end{eqnarray}	   
 \end{mathletters}   
 
 It should be noticed that the expressions derived above in the 
 framework
 of the Gaussian approximation for the 
 correlations $\widehat{C}^{(2)}_{G}$, and ${G}^{(2)}_{G,c}$, 
 and for the 
 free ennergy $f_{G}$ coincide with those obtained in the framework of 
 the RPA theory of simple fluids\cite{Hansen}. 
 As well known,  the RPA theory becomes exact in the limit of an infinitely
long-ranged interaction $w(r)=\gamma^3 w_{0}(\gamma r) , \; \gamma 
\to 0$ \cite{Hansen}.

 Note that, as in the 
 case of the MF theory, the free energy\ (\ref{freeG}) 
 depends on the value of the pair potential $w$ in the core, which is 
 arbitrary. We can take advantage of this freedom of the theory by 
 requiring the stationarity condition
 \begin{equation}
     \label{stationa}
 \frac{\delta \beta f_{G}}{\delta \beta w(\vec{r})} \equiv 0  
 \text{  for } \vert \vert \vec{r} \vert \vert \leq \sigma \; .
 \end{equation}
 For it is easy to derive from\ (\ref{freeG}) that
 $$\frac{\delta \beta f_{G}}{\delta \beta w(\vec{r})}   =
 -\frac{\rho^{2}}{2} g_{G}(\vec{r})\; ,$$ the stationarity 
 condition\ (\ref{stationa})
 on $\beta f_{G}$
 implies the nullity of the radial pair distribution function in the 
 core, an important physical requirement.
 This optimized Gaussian approximation coincides of course with 
 the ORPA theory of the theory of liquids\cite{Hansen}.
\section{The Landau-Ginzburg Action}
\label{LG}
We first note that, quite remarkably, all the results of the previous 
sections could have been obtained more simply by ignoring the complications 
introduced by the presence of  two distinct KSSHE fields $\varphi_{+}$ 
and $\varphi_{-}$ associated with the repulsive and attractive part 
of the pair potential respectively, and by considering rather a unique 
KSSHE field $\varphi$. Indeed, taking as a starting point the simplified
definitions
\begin{mathletters}
    \label{sloppy}
 \begin{eqnarray}
     \Xi[\nu] &= & {\cal N}_{\beta w}^{-1} \int {\cal D} 
     \varphi(\vec{r}) 
 \exp(-{\cal H}[\varphi]) \; , \\
  {\cal H}[\varphi] &=& \frac{1}{2}
  \left<\varphi \vert (\beta w)^{-1}\vert \varphi \right> -\log \Xi_{HS}
  [\overline{\nu}] \; , \\ 
  {\cal N}_{\beta w} &=&  \int {\cal D} \varphi(\vec{r})
  \; \exp \left( -  \left<\varphi \vert (\beta w)^{-1}\vert \varphi \right> \right)
  \end{eqnarray}  
 \end{mathletters}   
without any requirement on the positivity of the quadratic form
$<\varphi \vert (\beta w)^{-1}\vert \varphi>$,
yields the MF and Gaussian expressions for the MF and Gaussian 
expressions for $\log \Xi$, the free energy and the correlation 
functions. This remark justifies {\em a posteriori} the notations 
(\ref{caval}) adopted at the beginning of section\ \ref{KSSHE_Weight}. 
Henceforth, in order to avoid any awkward complications in further 
developments we shall adopt definitively the somehow abusive 
definitions\ (\ref{sloppy}).

We show now that the action ${\cal H}[\varphi]$ can be rewritten as the 
Landau-Ginzburg (LG) action of a magnetic system in the presence of a 
magnetic field $B$. Following Brilliantov \cite{Brilliantov} 
we make choice of a 
reference chemical potential $\nu_{0}$ for the HS reference system
(supposed to be uniform for simplicity and to be specified further) and 
perform a functional Taylor expansion of $\log \Xi_{HS}$ around
$\nu_{0}$. 
Let us first define for further convenience
\begin{mathletters}
    \label{defs}
    \begin{eqnarray}
\Delta \nu (\vec{r}) &=& \nu (\vec{r})- \nu_{S} - \nu_{0}  \; , \\
\phi (\vec{r}) &=& \Delta \nu (\vec{r}) + \varphi (\vec{r})  \; .
    \end{eqnarray}
\end{mathletters} 
With these definitions we have $\overline{\nu}=\nu_{0} +\phi$.
Now, assuming the analiticity of $\log \Xi_{HS}$ 
at $\nu_{0}$ 
we have, for $\overline{\nu}$ in some neighborhood of $\nu_{0}$ 
\begin{mathletters}
    \label{taylor}
    \begin{eqnarray}
  \log \Xi_{HS} [\overline{\nu}]  &=& \log \Xi_{HS} [\nu_{0}]   
  + \left< \phi \vert \rho_{HS}[\nu_{0}] \right > 
  +\frac{1}{2} \left< \phi \vert G_{HS,c}^{(2)}[\nu_{0}] \vert \phi
  \right >
  -{\cal V}_{I}[\phi] \; , \\
  {\cal V}_{I}[\phi] &=&-\sum_{n=3}^{\infty} \frac{1}{n!}
  \int d1 \ldots dn \; G_{HS,c}^{(n)}[\nu_{0}](1, \ldots ,n) \phi(1) 
  \ldots \phi(n) \; .
    \end{eqnarray}
\end{mathletters} 
Therefore one can rewrite the expression\ (\ref{H}) of $\Xi[\nu] $ as
\begin{mathletters}
    \label{LG1}
    \begin{eqnarray}
	\Xi[\nu] &=& 
	\exp( -\overline{{\cal H}}) \; \Xi_{LG}€[B] \; , \\
	\overline{{\cal H}}&=& \frac{1}{2}
	\left< \Delta \nu \vert (\beta w)^{-1} \vert 
   \Delta \nu 	\right > -\log
	\Xi_{HS}[\nu_{0}] \; ,
    \end{eqnarray}
\end{mathletters} 
where $\overline{{\cal H}}$ is a simple Gaussian functional of $\nu$ 
 and $\Xi_{LG}[B]$ may be 
seen as the partition function of a LG model in the presence of a 
magnetic field $B$. Indeed 
\begin{mathletters}
    \label{LG2}
    \begin{eqnarray}
\Xi_{LG}€[B]&=& {\cal N}_{\beta w}^{-1} \int {\cal D} 
     \phi(\vec{r}) 
 \exp(-{\cal H}_{LG,0}€[\phi] +\left< B \vert \phi \right>) \; , \\
 {\cal H}_{LG,0}€[\phi] &=& \frac{1}{2} \left< \phi \vert 
 \Delta^{-1}\vert \phi \right>  + {\cal V_{I}}[\phi] \; ,
    \end{eqnarray}
\end{mathletters} 
where the free propagator of the theory $\Delta$ is given by 
$\Delta^{-1}= (\beta w)^{-1}-G_{HS,c}^{(2)}[\nu_{0}]$ or equivalently
\begin{equation}
    \Delta=\beta w *(1-\beta w *G_{HS,c}^{(2)}[\nu_{0}])^{-1} \; ,
 \end{equation}
and the magnetic field $B$ by
\begin{equation}
    \label{B}
    B= \rho_{HS}[\nu_{0}] + (\beta w)^{-1}*\Delta \nu \; .
\end{equation}    
It can be noticed that the interaction term ${\cal V_{I}}[\phi]$ of 
the LG action 	${\cal H}_{LG,0}$ is non local and does not exhibit the 
usual symetry ${\cal V_{I}}[\phi] = {\cal V_{I}}[-\phi]$ of Isink like 
systems.
The correlation functions of this field theory are defined as usual 
as 
\begin{mathletters}
    \label{correphi}
    \begin{eqnarray}
 G_{\phi}^{(n)}[B](1,\ldots,n) &=&\left< \prod_{i=1}^{n}\phi(i)
 \right >_{{\cal H}_{LG}}	\nonumber \\
 &=& \frac{1}{ \Xi_{LG}} \frac{\delta^{n} \; \Xi_{LG}[B]}
 {\delta B(1) \ldots \delta B(n)} \; , \\
G_{\phi,c}^{(n)}[B] (1,\ldots,n)&=& \frac{\delta^{n} \; \log \Xi_{LG}[B]}
 {\delta B(1) \ldots \delta B(n)} \; .
    \end{eqnarray}
\end{mathletters} 

Since it follows from the definition\ (\ref{B}) of $B$ that
\begin{equation}
    \label{gyt}
    \frac{\delta}{\delta \nu}= (\beta w)^{-1} * \frac{\delta }{\delta 
    B} \; ,
\end{equation}
we deduce readily from the relation $\log \Xi [\nu] = \log
\Xi_{HS}[\nu_{0}] + \log \Xi_{LG}[B] -<\Delta \nu\vert (\beta 
w)^{-1}\vert \Delta \nu >/2$ that $\rho =(\beta 
w)^{-1}*(<\phi>_{{\cal H}_{LG}€}-\Delta \nu) \equiv (\beta 
w)^{-1}*<\varphi>_{{\cal H}}$, which coincides with relation\ 
(\ref{cc}). Performing then $n$ successive functional derivatives of 
$\log \Xi [\nu]$ in taking into account eq.\ (\ref{gyt}) one 
proves easily the following relations 
\begin{mathletters}
    \label{correphi2}
    \begin{eqnarray}
G_{c}^{(2)}[\nu] &=& -(\beta w)^{-1} + \overline{G}_{\phi,c}^{(2)}[B] 
\; \\
G_{c}^{(n)}[\nu] &=& \overline{G}_{\phi,c}^{(n)}[B]  \; \; \; \text{    for } 
n\geq 3 \; , 
    \end{eqnarray}
 where 
 \begin{equation}
 \overline{G}_{\phi,c}^{(n)}[B](1,\ldots,n) =
 \int d1' \ldots dn' (\beta w)^{-1}(1,1') \ldots (\beta w)^{-1}(n,n') \;
 G_{\phi,c}^{(n)}[B](1',\ldots,n') \; .
  \end{equation}   
\end{mathletters} 
Since the KSSHE fields $\phi$ and $\varphi$ differ by a constant 
$\Delta \nu$ the connected correlation functions $G_{\phi,c}^{(n)}[B]$ 
and $G_{\varphi,c}^{(n)}[\nu]$ are equal for $n\geq 2$. Note that in 
the case $n=2$ eq.\ (\ref{correphi2}) coincides fortunately with
eq.\ (\ref{GGphi2}).

When evaluated for a homogeneous field $\Phi=cte$ the LG Hamiltonian 
${\cal H}_{LG}[\phi]=V h_{LG}[\phi]$ where $h_{LG}[\phi]$
takes the familiar form \cite{Ma,Dowrick}
\begin{equation}
    \label{kol}
 h_{LG}[\phi]=\frac{1}{2} r_{0}\phi^{2} -B\phi+
 \sum_{n=3}^{\infty} u_{n}\phi^n \; ,
\end{equation}  
where $r_{0}=K-K_{0}$, with $K=1/(\beta \tilde{w}(0))$ proportional to 
the temperature and $K_{0}=\rho_{HS}^{(1)}(\nu_{0})$. In this 
case we have simply 
$B=\rho_{HS}(\nu_{0}) + K \Delta \nu$ and finally $u_{n}=-\beta 
P_{HS}^{(n)}(\nu_{0})=-\rho_{HS}^{(n-1)}(\nu_{0})$. Note that,
in general, odd and even powers of $\phi$ enter the expression of 
$h_{LG}[\phi]$. 
\section{The liquid-vapour transition}
\label{CP}
\subsection{The mean field KSSHE theory}
\subsubsection{General discussion}
\label{general}
We apply the results obtained on the previous sections  to discuss now
the liquid-vapour 
transition. The subject was discussed to some extent by 
Brillantov in ref. \cite{Brilliantov} with
however some inaccuracies that we correct and some 
missing points that we include in our discussion.

\begin{figure}
\begin{center}
\epsfxsize=5.9 truein
\epsfbox{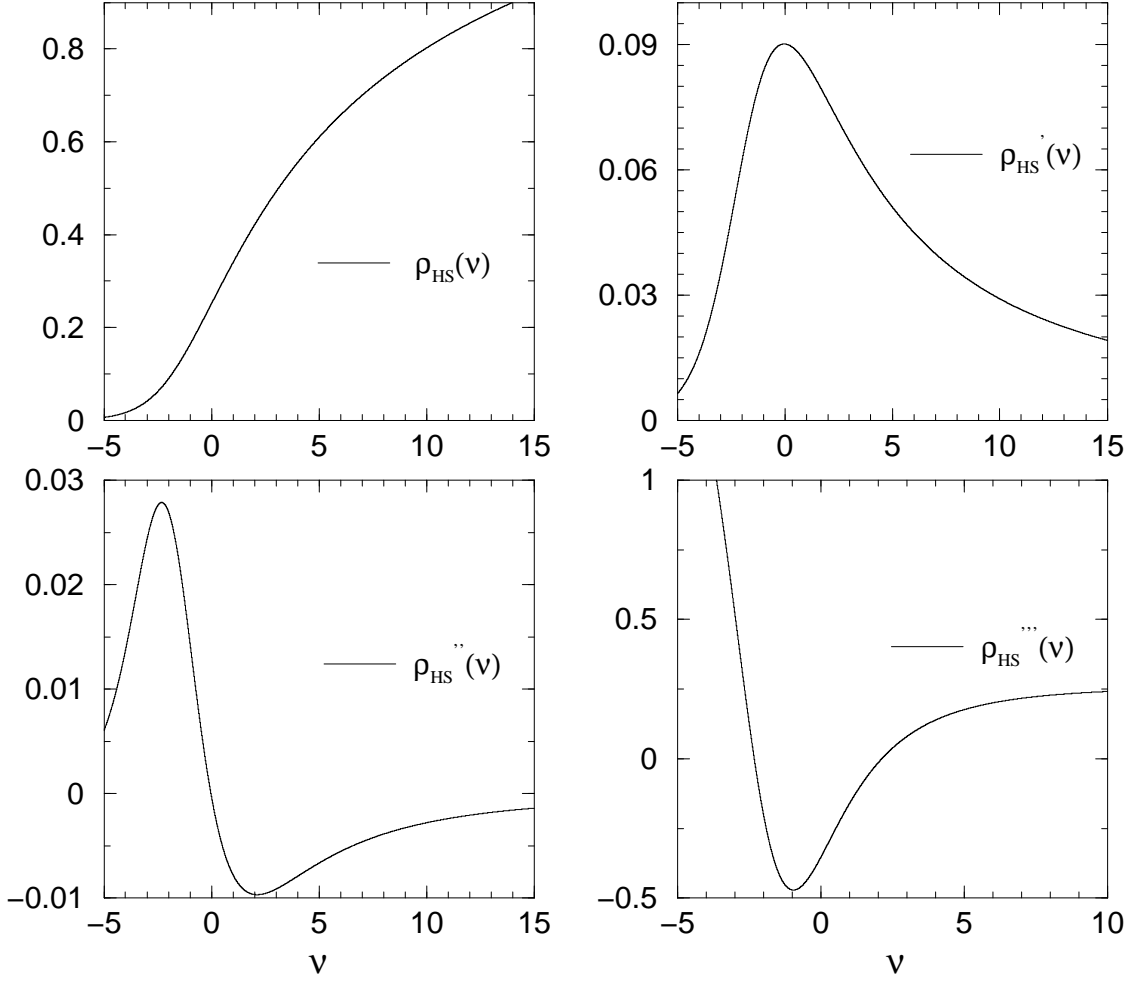}
 \caption{ The HS density $\rho_{HS}(\nu)$ and its first derivatives 
$\rho_{HS}^{(n)}(\nu)$ versus the chemical potential $\nu$.
The graphs were drawn with the help of  a parametric representation 
$[\rho_{HS}^{(n)}(\nu)=a(\eta), \nu=b(\eta) ]$ ($\eta$ 
packing fraction) extracted from the CS equation of state\cite{Hansen}.}
\label{fig1}
\end{center}
\end{figure}

With the notations of section\ \ref{LG} the MF equations\ 
(\ref{statio}) take the familiar form
\begin{equation}
    \label{LGB}
   \left. \frac{\delta {\cal H}_{LG,0}}{\delta \phi} \right 
   \vert_{\overline{\phi}}=B \equiv \rho_{HS}[\nu_{0}] + (\beta w)^{-1}
   *\Delta \nu
   \; .
 \end{equation}  
 Eq.\ (\ref{LGB}) gives the MF magnetisation of the  magnetic 
 system  associated to the fluid in the presence of the magnetic field $B$.
 For a homogeneous system we seek a solution 
 $\phi=\overline{\phi}=cte$ of eq.\ (\ref{LGB}). This is equivalent 
 to find the minima of the function $h(\phi)=h_{LG}(\phi) + K \Delta 
 \nu^{2}/2 - \beta P_{HS}(\nu_{0})$. We have
 \begin{mathletters}
     \label{eqs}
 \begin{eqnarray}
     \label{eqsa}
     h(\phi)&=&\frac{K}{2} \Delta \nu^{2} +\frac{K}{2}\phi^{2}
                  -K \Delta \nu \phi - \beta P_{HS}(\nu_{0}+\phi) 
                  \; ; \\
  \label{eqsb}  h'(\phi)&=&K \phi -K \Delta \nu -\rho_{HS}(\nu_{0} + \phi) \; , \\
    h''(\phi)&=& K-\rho_{HS}'(\nu_{0} + \phi) \; .
 \end{eqnarray}
 \end{mathletters}
 Moreover the link between the fluid and the associated magnetic 
 system is made through the relation
 \begin{equation}
    \label{Bhomo}
    B= \rho_{HS}(\nu_{0}) + K \Delta \nu \; .
 \end{equation}   
 Let us first discuss the case $K>0 \Longleftrightarrow 
 \widetilde{w}(0)>0$, i.e. when the pair potential is more attractive 
 than repulsive.
It is clear from eqs.\ (\ref{eqs}) that the convexity 
of the function $\beta P_{HS}(\nu)$ governs the 
properties of the function $h(\phi)$. Since $\beta P_{HS}(\nu)$ is 
convex its second derivative $\beta P_{HS}''(\nu)=\rho_{HS}'(\nu)$ is 
positive for all $\nu$. Moreover for all reasonable equation of state of the HS 
fluid the function $\beta P_{HS}''(\nu)$ exhibits a unique maximum at 
some $\nu_{0}=\nu_{0}^{*}$ as can be seen in figure (\ref{fig1})
where the graphs 
of some derivatives $\rho_{HS}^{(n)}(\nu)$ of the HS density are sketched 
within the framework of the Carnahan-Starling (CS) 
approximation\cite{Hansen}. In the 
CS approximation we have $\nu_{0}^{*}=-0.025$ and 
$\beta P_{HS}''(\nu_{0}^{*})=K_{0}=0.090$.
Similar values are found in the 
Percus-Yevick (PY) approximation \cite{Hansen} 
(via the compressibilty route which 
is simpler, see table\ I) since, at moderate densities, the predictions 
of the two theories are in close agreement. 

Therefore, for $K > K_{0}$ 
we have $h''(\phi) \geq 0 $ for all values of $\phi$ which implies 
that $h(\phi)$ is convex. Since $\lim_{\phi \to \pm 
\infty}h(\phi)=+\infty$ then $h(\phi)$ has one minimum for some
$\overline{\phi}$ and this minimum is unique. $\nu_{0}$ being given, 
let us choose now $\nu$ such that $B=0$ then it follows from eqs.\ 
(\ref{Bhomo}) and (\ref{eqsb}) that the MF equation 
$h'(\overline{\phi})=0$ can  be rewritten as
\begin{equation}
    \label{hh}
    K \overline{\phi}=\rho_{HS}(\nu_{0}+ \overline{\phi})
    -\rho_{HS}(\nu_{0}) \; .
 \end{equation}   
 A solution of eq.\ (\ref{hh}) is obviously $\overline{\phi}=0$ and 
 it is the unique solution, for $K > K_{0}$. For $K \leq K_{0}$ the 
 function $h(\phi)$ is no longer convex and other 
 solutions arise.
 
 Let us assume now that $K \leq K_{0}$ . We examine first the case 
 $\nu_{0}=\nu_{0}^{*}$. Noting for convenience $\phi^{*}$ the KSSHE 
 field and $h^{*}$ the Landau function 
 in this case we have from eq.\ (\ref{kol})
  \begin{mathletters}
     \label{eqs*}
 \begin{eqnarray}
     \label{eq*a}
     h^{*}(\phi^{*})&=&\frac{K}{2} \Delta \nu^{*2} +
       \frac{r_{0}}{2}\phi^{*2} -B^{*}\phi^{*} +\frac{u_{4}^{*}}{4!}
       \phi^{*4}
       +{\cal O}(\phi^{*5})
                  \; ; \\
   B^{*}&=&  \rho_{HS}(\nu_{0}^{*}) + K \Delta \nu^{*} \; , \\
   \Delta \nu^{*}  &=&\nu -\nu_{S} - \nu_{0}^{*}\; ,\\
   u_{4}^{*}&=& -  \rho_{HS}^{(3)} (\nu_{0}^{*}) >0  \; .
 \end{eqnarray}
 \end{mathletters}
 Since $u_{3}^{*}=-\rho_{HS}^{(2)} (\nu_{0}^{*})=0$ and moreover 
 $u_{4}^{*}\equiv - \rho_{HS}^{(3)} (\nu_{0}^{*})>0$
 (see the table and figure\ (\ref{fig1})) the Landau function $h^{*}(\phi^{*})$ 
 describes a $2^{nd}$ order phase transition with a  critical point at
 $(B^{*}=0,K=K_{0})$. For $K \leq K_{0}$ and $B^{*}=0$ the solution 
 $\overline{\phi}=0$ of eq.\ (\ref{hh}) becomes unstable and two 
 stable solutions $\pm \overline{\phi}_{0}\neq 0$  corresponding to a 
 liquid and a vapour phase emerge.
 
\begin{figure}
\epsfxsize=5.9 truein
\epsfbox{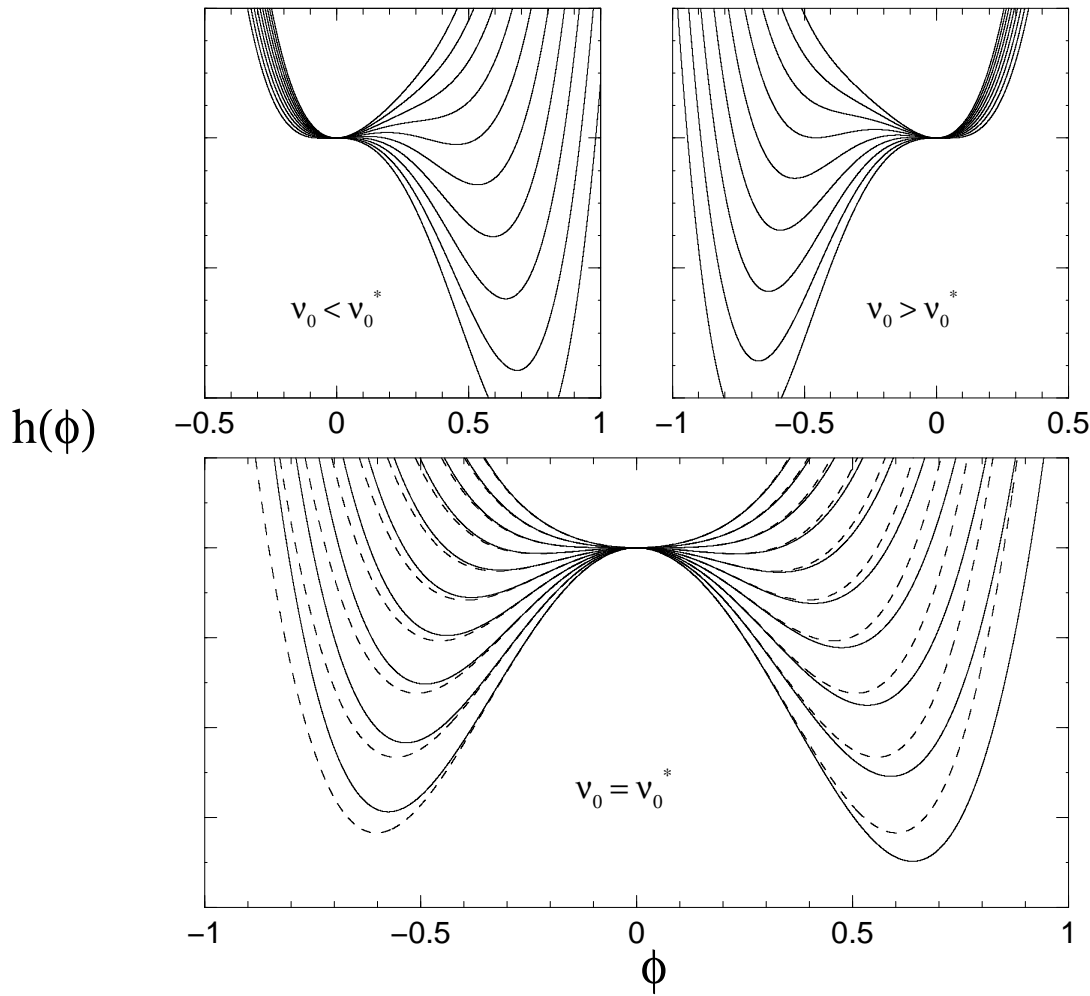}
\caption{ The Landau-Ginzburg Hamiltonian $h(\phi)$ in zero field 
($B=0$) for various choices of the reference chemical potential 
$\nu_{0}$ and for different temperatures $K$ in the vicinity of 
$K_{c}$. In each bunch of curves $K$ increases from bottom to top.
$h(\phi)$ was computed in the CS approximation.
Only the choice $\nu_{0}=\nu_{0}^{*}$ yields a second order phase 
transition with a critical point (bottom curve). In this case the 
dashed symmetric curves where obtained by truncating  $h(\phi)$ at 
the order $\phi^{4}$ (included). For $\nu_{0}\neq
\nu_{0}^{*}$, $h(\phi)$ describes a first order phase transition (top 
curves) without critical point.
The units of $h(\phi)$ were not specified for more clarity.}
\label{fig2}
\end{figure}

 Now if $\nu_{0}\neq \nu_{0}^{*}$ a term in $\phi^{3}$ is present in 
 the Landau function $h(\phi)$ at $B=0$ which describes now a first order 
 transition without critical point. This apparent paradox is solved in the 
 next section.  A sketch of the functions 
 $h^{*}(\phi^{*})$ and $h(\phi)$ is given in figure\ (\ref{fig2}) to 
 illustrate our discussion.
 
 Let us now consider briefly the case $K<0$ of a repulsive tail. In this 
 case $h^{(2)}(\phi)$ is negative for all $\phi$ and the function 
 $h(\phi), \; \phi \in \R $ is concave with a unique {\em maximum} for some
 $\overline{\phi}$. For $\phi$ assuming imaginary values 
 $h(\overline{\phi})$ is obviously a minimum of $h(\phi)$
 and the point $(\overline{\phi},h(\overline{\phi}))$ is indeed a 
 saddle point. Therefore for $K<0$ no transition occurs, at least at the MF 
 level. Henceforth we shall assume that $K>0$.
 \subsubsection{The Hubbard-Schofield transform and the critical point}
 When we make choice of the special value $\nu_{0}=\nu_{0}^{*}$ the 
 Landau function has the following exact expression
 \begin{equation}
     \label{jil}
     h^{*}(\phi^{*})= \frac{K}{2} \Delta \nu^{*2} +
     \frac{1}{2}\phi^{*2} -
     K \Delta \nu^{*} \phi^{*} -
     \beta P_{HS}(\nu_{0}^{*}+\phi^{*}) \; ,
  \end{equation}
  where we have kept the notations of section\ (\ref{general}). A Taylor 
  expansion of eq.\ (\ref{jil}) around $\phi^{*}=0 $ gives back the 
  usual ferromagnetic LG 
  Hamiltonian\ (\ref{eq*a}) with no term in $\phi^{*3}$.
  The following set of transformations
   \begin{mathletters}
     \label{hubb}
 \begin{eqnarray}
     \phi &=& \phi^{*}+\nu_{0}^{*}-\nu_{0} \; , \\
    \Delta \nu&=&\Delta \nu^{*} +\nu_{0}^{*}-\nu_{0} \; , 
 \end{eqnarray}
 \end{mathletters}
 obviously enables us to write $h(\phi)\equiv h^{*}(\phi^{*})$ where 
 $h(\phi)$ is given precisely by eq.\ (\ref{eqsa}). 
 The relations\ 
 (\ref{hubb}) constitute the exact Hubbard-Schofield transformation 
 which allows to get rid of the term in $\phi^{3}$ in
 $h(\phi)$\cite{Hubbard1}. Note 
 that under this transformation 
 \begin{equation}
     \label{hubbB}
     B= B^{*} +K(\nu_{0}^{*}-\nu_{0})+
     \rho_{HS}(\nu_{0} )-\rho_{HS}(\nu_{0}^{*}) \; ,
 \end{equation}
 as can be infered readily from the definition\ (\ref{Bhomo}) of the 
 magnetic field $B$. As a consequence, the function $h(\phi)$ at zero 
 field which describes a system undergoing 
 a {\em first order phase transition} may be seen as the LG 
 Hamiltonian associated with a system undergoing 
 a {\em second order phase transition} in presence of the field $ B^{*} =
 -K(\nu_{0}^{*}-\nu_{0})  - \rho_{HS}(\nu_{0} )
 +\rho_{HS}(\nu_{0}^{*})$. 
 
 Henceforth we shall adopt for definitness the choice 
 $\nu_{0}=\nu_{0}^{*}$ and drop all the subscripts "*" to simplify 
 the notations. With this choice, the liquid is associated with a 
 magnetic system with magnetisation $\phi$ 
 and the liquid-vapour transition and its 
 critical point can be made in a one to one correspondance with the 
 para-ferromagnetic (second order) transition and its  critical point.
 Since the magnetisation and the 
 density of the fluid are equal (up to a multiplicative constant) the 
 order parameter of the liquid-vapour transition is merely the 
 density. The field-mixing hypothesis of Rehr and 
 Mermin\cite{Rehr,Bruce} is therefore meaningless in the present 
 theory.
 
 The MF critical point of the liquid-vapour transition is therefore 
 defined by the conditions 
 \begin{mathletters}
     \begin{eqnarray}
	 \label{CPa}
 K_{c}&=&K_{0} \; , \\
                       \label{CPb}
 \overline{\phi}_{c}&=& 0  \; , \\
                       \label{CPc}
 B_{c}&=& 0\; .
 \end{eqnarray}
 \end{mathletters}
 The first condition\ (\ref{CPa}) can be satisfied only if $
 \widetilde{w}(0) \equiv (\widetilde{w}_{-}(0)- 
 \widetilde{w}_{+}(0))>0$, i.e. if the pair potential is more 
 attractive than repulsive. $K_{c}=\rho_{HS}^{(1)}(\nu_{0})
 \sim 0.290$ (see the table) 
 is a universal quantity in the sense that it does not depend on the 
 pair potential, however the critical temperature 
 $T_{c}=K_{c}/ \widetilde{w}(0) $ is not universal and depends in 
 particular upon the regularisation of $w(r)$ in the core. The second 
 condition\ (\ref{CPb}) determines the critical density since it 
 follows from\ (\ref{roMF2})
 that $\rho_{c}=\rho_{HS}(\nu_{0}) \sim 0.249$ (see the 
 table). $\rho_{c}$ is therefore also a universal quantity.  
 Finally the third
 condition\ (\ref{CPc}) determines the critical chemical potential 
 $\Delta \nu_{c}=-\rho_{HS}(\nu_{0})/\rho_{HS}^{(1)}(\nu_{0})$.
 $\Delta \nu_{c}$ is universal but not 
 $\nu_{c}$ which once again depends upon $w$ and its regularisation.
 
 All the conclusions concerning the universality of some of the critical 
 parameters in the MF approximation originate in our  choice of treating 
 at the same level the attractive and the repulsive part of the pair 
 interaction. If the repulsive part $w_{+}$ is treated exactly and 
 included in the reference system then $\rho_{c}$, $K_{c}$, and 
 $\Delta \nu_{c}$ are no longer universal quantities.
 
 In ref. \cite{Brilliantov} an
 approximate Hubbard-Schofield 
 transformation was devised by considering 
 the truncated Taylor expansion of $h(\phi)$ rather than 
 its exact expression\ (\ref{eqsa}); it yields 
  $\rho_{c}\neq \rho_{HS}(\nu_{0} )$ and that  $\rho_{c}$
 can moreover be adjusted at will to fit with
 experimental or numerical simulation results; these two conclusions 
 are however incorrect.

 \subsubsection{Mean Field solution for $T>T_{c}$}
 \label{hightc}
 We have seen in section\ (\ref{general}) that for $K > K_{c} \equiv 
 \rho^{(1)}_{HS}(\nu_{0})$,
 the MF solution $\overline{\phi}$ is unique and satifies the relation
 \begin{equation}
     \label{MFhight}
     h'(\overline{\phi})=0 
     \Longleftrightarrow K \overline{\phi}= K \Delta \nu + 
     \rho_{HS}(\nu_{0}+ \overline{\phi}) \; .
  \end{equation}  
The general results of section\ (\ref{MF}) can be specialised to the 
uniform case and yield 
\begin{mathletters}
     \begin{eqnarray}
	 \label{Tha}
 \beta P_{MF}(\nu)&=& \beta P_{HS}(\nu_{0} + \overline{\phi})
 -\frac{1}{2K}\rho_{HS}^{2}(\nu_{0} + \overline{\phi}) 
 \; , \\
                       \label{Thb}
 \rho_{MF}(\nu)&=&   \rho_{HS}(\nu_{0} + \overline{\phi})  \; , \\
                       \label{Thc}
 \beta f_{MF}(\rho) &=&\beta f_{HS}(\rho)   - \frac{\rho^{2}}
 {2} \beta \widetilde{w}(0)+ \frac{\rho}{2} \beta w(0) \; .
 \end{eqnarray}
 \end{mathletters}
 Note that the critical pressure 
 \begin{equation}
     \label{Pc}
 \beta P_{c}=\beta P_{HS}(\nu_{0})-
 \frac{\rho_{HS}^{2}(\nu_{0})}{2 \rho_{HS}^{(1)}(\nu_{0})}
\end{equation}
 is a universal quantity (one finds  for the critical compressibilty 
 factor $Z_{c}=\beta P_{c}/\rho_{c}  \sim 0.3589 (CS) \; , \; 0.3598 
 (PYC)$). Note also that eq.\ (\ref{Thb}) implies that the critical 
 isochore $\rho=\rho_{c}$ is defined by the condition $B=0$, since in 
 the case the (unique) solution of\ (\ref{MFhight}) is 
 $\overline{\phi}=0$.
 
  We complete the above results by establishing
  now the MF expressions for the 
  internal energy, the specific heat and the compressibility. The 
  excess internal energy by unit of volume $u_{ex}$ is given by
  \begin{eqnarray}
   \label{u}
   u_{ex}(\rho,\beta)&=& \frac{\partial}{\partial \beta} \left[ \beta 
   f_{MF}(\rho)-\beta f_{HS}(\rho) \right] \; , \nonumber \\
   &=&\frac{\rho}{2}w(0) - \frac{\rho^{2}}{2}\widetilde{w}(0) \; ,
  \end{eqnarray} 
  from which it follows that in the MF approximation 
  the excess specific heat vanishes above $T_{c}$ : 
  \begin{equation}
      \label{cvsup}
      C_{V,ex}= \left. \frac{\partial u_{ex}}{\partial \beta}\right 
      \vert_{\rho}\equiv 0 \; .
   \end{equation} 
 The isothermal compressibiliy $\chi_{T}=\beta \rho^{-2} 
 (\partial \rho / \partial \nu)_{\beta}$ can be rewritten as  
 \begin{equation}
     \chi_{T}=\frac{\beta}{\rho_{HS}^{2}(\nu_{0}
     + \overline{\phi})} \; \rho^{(1)}_{HS}(\nu_{0}
     + \overline{\phi}) \; \left. \frac{\partial \overline{\phi}}{\partial \nu} 
     \right \vert_{\beta}€
     \; .
 \end{equation}    
 For it follows from the MF equation\ (\ref{MFhight}) that
 \begin{equation}\label{util}
    \left. \frac{\partial \overline{\phi}}{\partial \nu}\right 
    \vert_{\beta}€= 
   \frac{K}{K-\rho_{HS}^{(1)}(\nu_{0} + \overline{\phi})}
 \end{equation} 
 we have finally
 \begin{equation}
     \label{kit}
     \chi_{T}= \frac{\beta}{\rho_{HS}^{2}(\nu_{0}
     + \overline{\phi})} \;  \frac{K\rho_{HS}^{(1)}(\nu_{0} + \overline{\phi}) }
     {K-\rho_{HS}^{(1)}(\nu_{0} + \overline{\phi})} \; .
 \end{equation} 
 Alternatively,  $\chi_{T}$ could have been
 obtained from the Fourier transform 
 at $k=0$ of the two-body correlation function, by making use of a 
 well-known expression of the theory of liquids \cite{Hansen}
\begin{eqnarray}
  \chi_{T} &=&   \frac{\beta}{\rho_{HS}^{2}(\nu_{0}
     + \overline{\phi})} \; \widetilde{G}_{MF,c}[\nu](k=0) \; , 
     \nonumber \\ 
     &=&\frac{\beta}{\rho_{HS}^{2}(\nu_{0}
     + \overline{\phi})} \; \frac{\widetilde{G}^{(2)}_{HS,c}
[\nu_{0}
     + \overline{\phi}](0)}
{1- K^{-1} \widetilde{G}^{(2)}_{HS,c}
[ \nu_{0}
     + \overline{\phi}](0) } \; ,
  \end{eqnarray}   
  where we made use of eq.\ (\ref{Grpa2}). Noting that 
  $\widetilde{G}^{(2)}_{HS,c}
[\nu_{0}
     + \overline{\phi}](0)=\rho^{(1)}_{HS}(\nu_{0}
     + \overline{\phi}) $ we are led back to eq.\ (\ref{kit}). Along 
     the critical isochore ($\rho=\rho_{c} \Longleftrightarrow 
     \overline{\phi}=0$) one has simply
  \begin{equation}
     \label{kitc}    
  \chi_{T}(\rho_{c}) =\frac{\beta}{\rho_{c}^{2}} \; \frac{K_{c}}{K-K_{c}} 
  \; ,
   \end{equation} 
   with a divergence for $K \to K_{c}+$ with a MF critical exponent 
   $\gamma=1$ as expected.
   
   The behavior of $\widetilde{G}_{MF,c}(k)$ for $k\to 0$ along the 
   critical isochore is also easy to study. As a consequence of 
   eqs.\ (\ref{G2}) and\ (\ref{Grpa2}) one has
 \begin{equation}
  \widetilde{G}_{MF,c}[\rho_{c}](k)= \frac{\rho_{c}}
  {1-\rho_{c}(\widetilde{c}_{HS}[\rho_{c}](k) +\beta \widetilde{w}(k))}
  \; ,
  \end{equation} 
  where $\widetilde{c_{HS}€}[\rho_{c}](k)$ is the Fourier transform of 
  the usual direct correlation function of the HS fluid at the 
  density $\rho_{c}=\rho_{HS}(\nu_{0})$. At low $k$'s one has
  \begin{eqnarray}
     \widetilde{c}_{HS}[\rho_{c}](k) +\beta \widetilde{w}(k))&=
&   \widetilde{c}_{HS}[\rho_{c}](0) +\beta \widetilde{w}(0))  - a 
   k^{2} +{\cal O}( k^{4} ) \; , \nonumber \\
   \label{ae}
   a&=& \frac{-1}{2} [\beta \widetilde{w}^{(2)} (0)+
   \widetilde{c}_{HS}^{(2)}[\rho_{c}](0)] \; , 
  \end{eqnarray}
 which implies, putting all things together and after having noted that $1- 
 \rho_{c}\widetilde{c}_{HS}[\rho_{c}](0)=\rho_{c}/K_{0}$ that
 \begin{equation}
     \label{ytu}
     \widetilde{G}_{MF,c}[\rho_{c}](k) \sim \frac{K K_{0}}{K- K_{0}} \;
     \frac{1}{1+\xi^{2} k^{2}} \; \text{ for } k \to 0 \; ,
   \end{equation}  
 where the correlation length $\xi$ reads as 
 \begin{equation}
     \xi=\frac{a^{1/2}}{(K- K_{0})^{1/2}}
     \; .
 \end{equation}    
 Of course $\xi$ is defined only if $a>0$ which put some restrictions 
 on the pair potential (cf eq.\ (\ref{ae}) and note that
 $\widetilde{c}_{HS}^{(2)}[\rho_{c}](0)\sim 1.331 $ (PYC) ).
 $\xi$ diverges 
 for $K\to K_{c}+$ along the critical isochore 
 with, as expected, a MF exponent $\nu=1/2$. At the critical temperature 
 eq.\ (\ref{ytu}) implies that  $\widetilde{G}_{MF,c}(k) \sim k^{-2}$ 
 the usual behavior in the MF approximation yielding $\eta=0$ for 
 the Fisher exponent.
 
 We end this section by a discution of the behaviour of the 
 order parameter along the critical isotherm $K=K_{c}$. Near the 
 critical point the MF eq.\ (\ref{hightc}) at $T_{c}$
 can be reexpressed as
 \begin{eqnarray}
\overline{\phi}&=& \Delta \nu + \frac{\rho_{HS}(\nu_{0}+\overline{\phi} )}
{\rho_{HS}^{(1)}(\nu_{0})}  \; , \nonumber \\
&=&\Delta \nu - \Delta \nu_{c} + \overline{\phi} +
\frac{1}{3!} \frac{\rho_{HS}^{(3)}(\nu_{0})}{\rho_{HS}^{(1)}(\nu_{0})}
\overline{\phi}^{3} + {\cal O} (\overline{\phi}^{4} ) \; ,
\end{eqnarray} 
yielding 
 \begin{equation}
   \overline{\phi}^{3}= \frac{6}{u_{4}} (\nu- \nu_{c}) 
   \rho_{HS}^{(1)}(\nu_{0}) \; .
 \end{equation}
 Therefore along the critical isotherm we have finally
 \begin{equation}
     \rho = \rho_{HS}(\nu_{0}+  \overline{\phi}) \sim
     \rho_{HS}(\nu_{0}) + \rho_{HS}^{(1)}(\nu_{0})^{4/3}
     (6/u_{4})^{1/3} (\nu -\nu_{c})^{1/3} \;
  \end{equation}  
  yielding the classical value $\delta=3$ of the critical exponent, 
  as expected.
 \subsubsection{Mean Field solution for $T<T_{c}$}
 \label{lowtc} 
Below (and near) $T_{c}$ it is sufficient to consider the truncated 
expansion\ (\ref{eqs}) of the Landau function $h(\phi)$. The solution 
of the MF equation is of course well known in this case. Let us 
define $\widehat{B}=B/\vert B \vert $. Besides $\overline{\phi}=0$, the 
MF equations $h^{(1)}(\phi)=0$ have the 
solution\cite{Ma,Parisi,Dowrick}
\begin{eqnarray}
\overline{\phi}&=& \overline{\phi}_{0} \widehat{B}
+ \delta \overline{\phi} \; , 
\nonumber \\
\overline{\phi}_{0} &=& \left( \frac{3!}{u_{4}}\right)^{1/2}
\; (K_{0}- K)^{1/2}  \; , 
\nonumber \\ 
\delta \overline{\phi}&=& \frac{B}{2(K-K_{0})} + {\cal O}(B^{2}) \; .
\end{eqnarray}  
In zero field (i.e. $B=0$) the solution $\overline{\phi}=0$ is 
unstable and the two solutions $\overline{\phi}=\pm \vert \overline
{\phi}_{0} \vert$ are stable. The densities corresponding to the 
magnetisations 
$\pm \vert \overline{\phi}_{0} \vert$ are those of the coexisting 
liquid and vapour; they are given by
\begin{eqnarray}
    \label{densi}
    \rho_{l,g}&=& \rho_{HS}(\nu_{0}\pm \vert \overline
{\phi}_{0} \vert ) \; , \nonumber \\
&\sim&\rho_{HS}(\nu_{0}) \pm \rho_{HS}^{(1)}(\nu_{0}) 
\left( \frac{3!}{u_{4}}\right)^{1/2}
\; (K_{0}- K)^{1/2} \; ,
\end{eqnarray}
yielding a value $\beta = 1/2$ for the exponent of the order 
parameter, as expected. It can be noticed that eq.\ (\ref{densi}) 
supports the law of rectiligned diameters with the simple result 
(valid in the vicinity of $T_{c}$) $\rho_{l}+\rho_{g}=2 \times 
\rho_{c}$.
The chemical potential at the coexistence is given by $\Delta 
\nu_{coex}=-\rho_{HS}(\nu_{0})/K$ as a consequence of the condition 
$B=0$ and the pressure at coexistence reads as
\begin{equation}
    \beta P_{coex}= \beta P_{HS}(\nu_{0})-\frac{\rho_{c}^{2}}{2 K}
    +\frac{3}{2 u_{4}} (K-K_{0})^{2} \; 
 \end{equation}
 Therefore $\nu_{coex}$ and  $\beta P_{coex}$ are both regular 
 functions of the temperature $K$ near $K_{c}$.
 
 The compressibility is not defined in the two phases region but one 
 can compute the excess specific heat at constant volume thanks to a 
 formula due to Yang and Yang \cite{Yang}. Along the critical isochore one has
 \begin{equation}
  C_{V,ex}=T \frac{\partial^{2} P_{coex}}{\partial T^{2}}- 
  \rho_{c}T \frac{\partial^{2} \mu_{coex}}{\partial T^{2}} \; ,
  \end{equation}
 which yields after some algebra to
 \begin{equation}
     \label{cvinf}
  C_{V,ex}=\frac{3\rho_{HS}^{(1)}(\nu_{0})^2 }{2 u_{4}} 
  \ \; .
   \end{equation}
 The comparison of eqs.\ (\ref{cvsup}) and\  (\ref{cvinf}) show that 
 the specific heat at $\rho_{c}$ has a discontinuity of universal 
 value $\Delta C_{V,ex}\sim 3\rho_{HS}^{(1)}(\nu_{0})^2 
 /2 u_{4} \sim 1.0496 \; \text{(CS)}\; \text{ or }  1.0074 \; \text{ 
 (PYC)}$,
 yielding a MF value 
 $\alpha=0$ for the critical exponent of $C_{V,ex}$ as expected.
 \subsubsection{The Gaussian approximation}
 As well known, the critical exponents of the Gaussian model are the 
 same as those of the MF theory except $\alpha$. We shall thus content 
 ourselves to compute $C_{V,ex}$ along the critical isochore above 
 $T_{c}$. Since  $C_{V,ex}(\rho_{c})$ vanishes in the MF 
 approximation above $T_{c}$, it is equal to 
 $(\partial^{2} \beta \Delta f/ \partial 
 \beta^{2})_{\rho_{c}}$ in the Gaussian approximation where 
 $\beta \Delta f$ has been defined in eq.\ (\ref{freeGc}).
  Working out the derivatives one find that
 \begin{equation}
     \label{Cvgaussien}
  C_{V,ex}(\rho_{c})=\frac{1}{2}  \int \frac{d^{3}\vec{k}}{(2\pi)^{3}} \; 
 \left[ \widetilde{G}_{MF,c}^{(2)}[\rho_{c}](k) \;
  \beta \widetilde{w}(k) \right]^{2} \; .
   \end{equation}   
 The integral diverges in the limit $K\to K_{c}+$ because of the 
 singular behavior of $ \widetilde{G}_{MF,c}^{(2)}[\rho_{c}](k)$ at 
 small k's. The method for  extracting the divergence of 
 $ C_{V,ex}(\rho_{c})$ is well documented \cite{Ma,Parisi,Dowrick}
 and will not be repeated 
 here. The idea is to make the 
 change of variable $\vec{k}'=\xi \vec{k}$, where $\xi$ is the 
 correlation length and to inject in eq.\ 
 (\ref{Cvgaussien}) the small "k" behaviour of 
 $\widetilde{G}_{HS,c}^{(2)}[\rho_{c}](k)$ as given by eq.\ (\ref{ytu}).
 One finds for $K \to K_{c}+$
 \begin{equation}
     \label{Cvgaussien2}
    C_{V,ex}(\rho_{c})\sim \frac{\rho_{HS}^{(1)}(\nu_{0})^{2}}
    {16 \pi a^{3/2}} \; (K - K_{c})^{-1/2} \; , 
  \end{equation}      
 yielding a divergence in $t^{-1/2}$ as expected. Below $T_{c}$ a 
 similar behavior (with a different prefactor) can be obtained by 
 making use of the Yang-Yang formula. As well known,
 a comparison of the discontinuity $\Delta C_{V,ex}(\rho_{c})$ of the 
 specific heat in the MF approximation
 (cf eq.\ (\ref{cvinf})) and its Gaussian behavior \ (\ref{Cvgaussien2})
 allows an estimate of 
 the Ginzburg temperature range $\Delta K$, in our case one has
 $\Delta K \sim a^{3/2}/u_{4}$. It depends on the pair potential $w$ 
 through the parameter $a$ defined at eq.\ (\ref{ae}). Since $a\sim 
 1$ and $u_{4}$ is tiny in general, $\Delta K/K_{c} >1$.
 \section{CONCLUSION}
 \label{CONCLU}
 In this paper we have presented an exact field theoretical 
 representation of the statistical mechanics of a simple model of 
 liquid. The action of the theory is obtained with the help of a
 KSSHE transform of the Boltzman factor of the fluid. 
 Some complications of the formalism such that the fact that the 
 action of the KSSHE theory is not a local functional of the field 
 have their origin in the necessity to 
 take an explicit account of the hard core interactions. However, it 
 is possible to establish
 the relations between the correlation functions of the field and 
 that of the density of the fluid.  Moreover the KSSHE 
 action can be interpretated as the Landau-Ginzburg action of a 
 ferromagnetic system in the presence of a magnetic field $B$ which is
 related linearly with the local chemical potential of the liquid.
 
 The MF and Gaussian approximation of the KSSHE theory can be 
 worked out explicitly. For homogeneous fluids the Gaussian 
 approximation coincides with the RPA theory.
 These two approximations yield a rough description of the 
 liquid-vapour transition. The exact 
 Hubbard-Schofield transformation can be establihed 
 which yields
 the usual LG Hamiltonian in $\phi^{4}$ of the theory of critical 
 phenomena.  The density 
 emerges as the order parameter of the transition which rules out the 
 field mixing hypothesis.  The 
 critical density turns out to be independant from
 the details of the attractive part of the 
 potential which is a serious flaw. 
 Moreover several  critical properties either do not 
 depend on the potential $w(r)$ or depend 
 strongly on the regularisation of $w(r)$ in the core which is in both 
 cases unsatisfactory. By lack 
 of place we did not discuss the full one-loop order theory where 
 fresh difficulties arise such as unavoidable 
 corrections to the magnetic field $B$, the order parameter, and more 
 generally all the correlation functions.
 A fully renormalized theory at the one-loop
 order seems however feasible despite these difficulties.
 
 The MF-KSSHE theory also works for inhomogeneous fluids and yields 
 density functionals which could be usefull in applications. Moreover 
 for two classes of pair potentials, the MF-KSSHE free energy functional
 constitutes a  rigorous 
 bound for  the exact free energy functional which could serve as a test in 
 numerical studies. A similar result holds also for the 
 grand-potential functional above the critical point.
\acknowledgments
I acknowledge  D. Levesque
for useful discussions and comments. 
\appendix
\section{Functional Integration}
\label{a}

In the case where the domain occupied by the
particles of the fluid is a cube of side $L$ with periodic boundary conditions
( $\Omega \equiv {\cal C}_3$ ) we give an explicit expression of 
the measure 
${\cal D } \varphi $ which enters the functional
integrals considered in this paper.

In cubico-periodical geometries the microscopic density $\widehat{\rho}$ can be
written as a Fourier series

\begin{eqnarray}
\label{ap1}
\widehat{\rho}(\vec{r})&=&\frac{1}{L^3} 
\sum_{\vec{k} \in \Lambda} \widehat{\rho}_{\vec{k}} \; \exp(i \vec{k} .
\vec{r}) 
\; , \nonumber \\
\widehat{\rho}_{\vec{k}} &=& \sum_{i=1}^N \exp(-i \vec{k} .\vec{r}_i) \; ,
\end{eqnarray}
where $\Lambda = (2\pi/L) \; \Z^{3}$ denotes the reciprocal lattice. Note 
that since $\widehat{\rho}$ is real one has
\begin{equation}
\label{ap3}
\widehat{\rho}_{\vec{k}}=\widehat{\rho}_{-\vec{k}}^* \; .
\end{equation}
 
The pair
potentials $w^{{\cal C}^3}_{\pm}(\vec{r}_{ij})$ 
between two particles $i$ and $j$ must take into account the 
interactions between the
periodical images of $i$ and $j$. They are both  periodical functions of
$\vec{r}_{ij}$ which can be written as \cite{Hansen} 

\begin{equation}
\label{ap2}
w^{{\cal C}^3}_{\pm}(\vec{r}) = \sum_{\vec{n} \in \Z^3}
 w_{\pm}(\vec{r}+L\vec{n}) \; ,
\end{equation}
The above expression for $w^{{\cal C}^3}_{\pm}(\vec{r})$ 
can also be written as a Fourier series,
the Fourier coefficients of which
are precisely the Fourier transforms of $w_{\pm}(r)$ as a consequence of Poisson's
formula. Therefore one also has 
\begin{equation}
w^{{\cal C}^3}_{\pm}(\vec{r})=\frac{1}{L^3} \sum_{\vec{k} \in \Lambda} 
 \widetilde{w}_{\pm}(k) \exp(i \vec{k} .\vec{r}) \; .
\end{equation}
where $\widetilde{w}_{\pm}(k)$ is real and has been defined at eq.\ (\ref{w}).
It follows from eqs\ (\ref{ap1})
and\ (\ref{ap2}) that, in ${\cal C}^3$,
the potential energies  $\left< \widehat{\rho}^{\;{\cal C}^3} \; 
 \vert w_{\pm}^{\;{\cal C}^3}\;  \vert
\widehat{\rho}^{\;{\cal C}^3}\; \right>$ in ${\cal C}^3$
can be reexpressed as
\begin{equation}
\frac{1}{2} \left< \widehat{\rho}^{\;{\cal C}^3} \;  \vert w_{\pm}^{\;{\cal C}^3}
\;  \vert
\widehat{\rho}^{\;{\cal C}^3} \; \right> = \frac{1}{L^3}
\sum_{\vec{k} \in \Lambda^{*}} \widetilde{w}_{\pm}(k) 
\vert \widehat{\rho}_{\vec{k}}
\vert ^{2} \; ,
\end{equation}
where the sum in the r.h.s runs over only the half $\Lambda^{*}$ of
all the vectors of the reciprocal lattice
$\Lambda$ (for instance those with $n_x \geq 0$) as a consequence 
of the reality of 
$\widetilde{w}_{\pm}(k)$ and of the symetry relations\ (\ref{ap3}).
 
At this point we introduce a periodic real field $\varphi(\vec{r})$ supposed to
be expressible as a Fourier series
\begin{equation}
\varphi(\vec{r})=\frac{1}{L^3} \sum_{\vec{k} \in \Lambda} \widetilde{\varphi}
(\vec{k}) \exp(i \vec{k} .\vec{r}) \; ,
\end{equation}
and define the measure \cite{Dowrick}
\begin{eqnarray}
\label{ap7}
{\cal D } \varphi & \equiv&  \prod_{\vec{k} \in \Lambda^{*}} d^2
 \widetilde{\varphi}(\vec{k}) \;, \nonumber \\
 d^2
 \widetilde{\varphi}(\vec{k}) &=& d\;\Re{\widetilde{\varphi}(\vec{k})}
 \;  d\;\Im{\widetilde{\varphi}(\vec{k})} \; ,
 \end{eqnarray}
where it is understood that the domain of integration of both
the real and imaginary
parts of $\widetilde{\varphi}(\vec{k})$ is the whole real axis
$]-\infty,+\infty[$. 

\noindent Defining formally 
\begin{eqnarray}
\label{ap8}
\left< \varphi \vert (\beta w_{\pm}^{{\cal C}_3})^{-1} \vert \varphi \right >
& =& 
\frac{1}{L^3} \sum_{\vec{k} \in \Lambda} \frac{1}{\beta \widetilde{w}_{\pm}(k)}
\vert \widetilde{\varphi}
(\vec{k})\vert^2 \; , \nonumber \\
& =&  \frac{1}{L^3} \sum_{\vec{k} \in \Lambda}
 \frac{1}{\beta \widetilde{w}_{\pm}(k)}\left( 
 \Re{\widetilde{\varphi}(\vec{k})}^2 + \Im{\widetilde{\varphi}(\vec{k})}^2
 \right) \; ,
\end{eqnarray}
and noting that 
\begin{equation}
\left< \widehat{\rho} \vert \varphi \right> =
\frac{2}{L^3} 
\sum_{\vec{k} \in \Lambda^{*}}\left( 
\Re{\widetilde{\varphi}(\vec{k})} \Re\widehat{\rho}_{\vec{k}}
 \; + \; \Im{\widetilde{\varphi}(\vec{k})} \Im\widehat{\rho}_{\vec{k}}
 \right) \; ,
\end{equation}
one deduces easily from the definition\ (\ref{ap7}) and from
properties of ordinary Gaussian integrals
\begin{equation}
    \int_{-\infty}^{+\infty} dx \; \exp(-\frac{1}{2} A x^2 +i J x)
    = (2\pi/A)^{1/2} \exp(-\frac{1}{2} A^{-1} J^2 ) \; (A >0) \; ,
\end{equation}   
the fundamental relations
\begin{eqnarray}
\label{fondap}
\exp\left(\frac{1}{2}  \left< \widehat{\rho} \vert
 \beta w_{-}^{{\cal C}_3} \vert \widehat{\rho} \right>  \right) &=& 
 \frac{\int {\cal D }\varphi \;
  \exp \left(  -\frac{1}{2}
  \left< \varphi \vert
  (\beta w_{-}^{{\cal C}_3})^{-1} \vert \varphi \right > +
  \left< \widehat{\rho} \vert \varphi \right> \right)}
  {\int {\cal D }\varphi \;
  \exp \left( -\frac{1}{2} \left< \varphi \vert
  (\beta w_{-}^{{\cal C}_3})^{-1} \vert \varphi \right >\right)}
 \equiv
 \left< \exp\left(\left< \widehat{\rho} \vert \varphi \right> 
\right)\right>_{\beta w_{-}^{{\cal C}_3}} 
 \; ,
 \nonumber \\
 \exp\left(-\frac{1}{2} \left< \widehat{\rho} \vert
 \beta w_{+}^{{\cal C}_3} \vert \widehat{\rho} \right> \right)&=& 
 \frac{\int {\cal D }\varphi \;
  \exp \left(  -\frac{1}{2}
  \left< \varphi \vert
  (\beta w_{+}^{{\cal C}_3})^{-1} \vert \varphi \right > + i 
  \left< \widehat{\rho} \vert \varphi \right> \right)}
  {\int {\cal D }\varphi \;
  \exp \left( -\frac{1}{2} \left< \varphi \vert
  (\beta w_{+}^{{\cal C}_3})^{-1} \vert \varphi \right >\right)}
 \equiv
 \left< \exp\left(  i \left < \widehat{\rho} \vert \varphi \right> 
\right)\right>_{\beta w_{+}^{{\cal C}_3}}   
 \; ,
\end{eqnarray}
already mentioned in the main text (cf eqs.\ (\ref{KSSHE1})) with 
less awkward notations.
An important result concerns the 
ratio of two Gaussian partition functions. Let ${\cal N}_{f}$ be the 
constant
\begin{equation}
    \label{norma}
 {\cal N}_{f}\equiv  \int {\cal D }\varphi \;
  \exp \left( -\frac{1}{2} \left< \varphi \vert
  f^{-1} \vert \varphi \right >\right) \; , 
\end{equation}
where $f$ is some positive two-body operator. Then, for $f$ and $g$ 
positive we have
\begin{eqnarray}
  {\cal N}_{f}/{\cal N}_{g}&=&\prod_{\vec{k} \in \Lambda^{*}}
  \frac{\widetilde{f}(\vec{k})}{\widetilde{g}(\vec{k})} \; , 
  \nonumber \\
  &\sim & \exp\left( \frac{V}{2} \int \frac{d^3 \vec{k}}{(2 \pi)^3}
  \log \frac{\widetilde{f}(\vec{k})}{\widetilde{g}(\vec{k})} \right) ; .
\end{eqnarray}  
In practice only functional integrals of Gaussian functionals
and their derivatives can be
performed explicitly. All the feasible integrals 
can be deduced from Wick's theorem 
\cite{Ma,Parisi,Zinn}
which states that
\begin{equation}
    \label{Wick}
\left< \varphi(1) \ldots \varphi(2n)\right>_{y}= \sum y(i_1,i_2) \ldots
y(i_{2n-1},i_{2n}) \; ,
\end{equation}
where the summation runs over all  the $(2n-1)!!$ distinct  pairs 
$(i_1,i_2)$. 

\newpage
\begin{table}
\caption{Derivatives of the HS pressure  at the chemical potential 
$\nu_{0}$ where $\beta P_{HS}^{(2)}$ 
reaches its maximum. The numerical estimates were obtained in the 
framework of the CS and PYC 
aproximations.}
\label{table}
 \begin{tabular}{|r|r|r|} 
 $       $   &      $   CS   $    &     $   PYC  $        \\ \hline
 $ \nu_{0}$    & $-0.025 $ & $-0.059 $     \\  \hline
 $ \beta P_{HS}(\nu_{0})$    & $0.433 $ & $0.425 $    \\ \hline
  $ \rho_{HS}(\nu_{0}) \equiv  \beta P_{HS}^{(1)}(\nu_{0})$  
  & $0.249$ & $0.246$     \\ \hline
 $  \rho_{HS}^{(1)}(\nu_{0})$    & $0.090$ & $0.0896$   \\ \hline
 $  \rho_{HS}^{(2)}(\nu_{0})$    & $0.0$ & $0.0$  \\ \hline
 $  u^{4} \equiv - \rho_{HS}^{(3)}(\nu_{0})$    & $0.0116$ & $0.0119$\\
 \hline
 $  \Delta \nu_{c}\equiv - \rho_{HS}(\nu_{0})/
 \rho_{HS}^{(1)}(\nu_{0})$    & $-2.766 $ & $-2.743$
\end{tabular}
\end{table}

\end{document}